\documentclass[draftclassnofoot,comsoc,doublecolumn, 10pt]{IEEEtran}
\UseRawInputEncoding
\usepackage[T1]{fontenc}
\usepackage{amsmath}
\usepackage[cmintegrals]{newtxmath}
\usepackage{graphicx}
\usepackage{color} 
\usepackage{epsf}
\usepackage{amsmath, amsfonts}
\usepackage{bbm}
\usepackage{latexsym}
\usepackage{subfigure}
\usepackage{multirow}
\usepackage{multicol}
\usepackage{cite}
\usepackage{float}
\usepackage{algorithm}
\usepackage{algpseudocode}
\usepackage{lipsum}

\newtheorem{lemma}{Lemma}

\newtheorem{theorem}{Theorem}
\newtheorem{proposition}{Proposition}

\newtheorem{definition}{Definition}

\newcommand{\argmax}{\operatornamewithlimits{argmax}}

\ifCLASSINFOpdf
\else
\fi
\usepackage{amsmath}
\usepackage[cmintegrals]{newtxmath}
\begin{document}
	
	\title{Optimal Scheduling Policy for Minimizing Age of Information with a Relay}
	
	\author{Jaeyoung~Song, ~\IEEEmembership{Member, ~IEEE}, Deniz Gunduz, ~\IEEEmembership{Fellow,~IEEE}, and Wan Choi, ~\IEEEmembership{Fellow,~IEEE}
		\thanks{This work was supported in part by the National Research Foundation of Korea (NRF) Grant funded by the Korea Government through the Ministry of Science and ICT (MSIT), South Korea, under Grant RS-2023-00220985. This work was supported in part by the National Research Foundation of Korea(NRF) grant funded by the Korea government(MSIT) (No. RS-2022-00166740).}
		\thanks{J. Song is with Department of Electronics Engineering, Pusan National University (PNU), Busan 46241, Korea (e-mail: jsong@pnu.edu)}
		\thanks{D. Gunduz is with the Department of Electrical and Electronic Engineering, Imperial College London, London SW7 2AZ, U.K. (e-mail: d.gunduz@imperial.ac.uk).}
		\thanks{W. Choi is with the Department of Electrical and Computer Engineering and the Institute of New Media and Communications, Seoul National University (SNU), Seoul 08826, Korea (email: wanchoi@snu.ac.kr).}    }
	
	
	\maketitle
	
	\begin{abstract}
		We investigate age of information in an Internet-of-things (IoT) sensor network where a single relay terminal connects multiple IoT sensors to their corresponding destination nodes. In order to minimize average weighted sum AoI, joint optimization of sampling and updating policy of a relay is studied. For error-free and symmetric case where weights are identical,  the necessary and sufficient condition for optimal policy is figured out. We also obtain the minimum average sum AoI in a closed-form expression which can be interpreted as the fundamental limit of sum AoI in a single relay network. Moreover, we  prove that the greedy policy is optimal for minimizing the average sum AoI at the destination nodes in the error-prone symmetric network. For general case where weights are arbitrarily given, we  propose a scheduling policy obtained via deep reinforcement learning.
	\end{abstract}
	
	\begin{IEEEkeywords}
		Age of information, Internet-of-things, Wireless sensor networks, Scheduling policy, Deep reinforcement learning
	\end{IEEEkeywords}
	
	\IEEEpeerreviewmaketitle
	\section{Introduction}\label{sec:introduction}
	With the rapid proliferation of Internet-of-things (IoT) applications, unprecedented volume of information is collected at the network edge. However, not every piece of information is equally valuable. In addition to the relevance of information for the underlying objective, in many emerging IoT applications, the value of information also depends on its timeliness. For example, in a sensor network, where sensors observe physical processes, once information about the status of the underlying process at a specific time is generated, the content of this observed information remains fixed while the corresponding process keeps changing over time. Therefore, for a time-varying process, it is expected that status information acquired more recently is more accurate than the one obtained in the past. Thus, the value of information depends highly on the time of generation and utilization.
	
	As a metric of assigning temporal values to information, age of information (AoI) has been introduced and investigated in many different areas \cite{Yates_Arxiv2020, Kosta_FTN2017, Kaul_ICSCSM2011, Klügel_INFOCOM2019, Krikidis_WCL2019, Tsai_INFOCOM2020}. AoI is defined as the time elapsed since the generation of information. Thus, AoI is a time-varying measure that takes into account   the crucial factors that may impact the timely delivery of information to the point of interest, including sampling, processing and delivery. While AoI is related to delay on a performance measure, delay focuses on the duration of the delivery, and ignores the freshness of the information available at the transmitter, which is critical to AoI. Therefore, minimizing the AoI requires new strategies that go beyond simply reducing the communication delay.
	
	Thanks to its fundamental nature and relevance in many applications, ranging from IoT to vehicular networks, AoI has gained significant popularity. In the context of IoT sensor networks, \cite{Arafa_TIT2020} explored AoI with energy harvesting. This research proposed average AoI and a new status transmission policy considering factors such as battery size and energy arrival process.  In \cite{Kaul_SMACN2011}, AoI was analyzed for vehicular networks. Since vehicular communication employs a random access protocol where transmission is decided with a probability, the study demonstrated that a solution that maximizes throughput cannot be directly applied to minimize the age of information in a random access channel. Additionally, an edge caching system was examined in \cite{Zhang_TWC2021}, focusing on a mobile edge caching system where content updates were considered. AoI was utilized to measure the timeliness of contents, and an AoI-aware content updating scheme was proposed and optimized. Moreover, the study explored the period for content updating.
	
	The concept of AoI has been extensively studied in scenarios where information is directly transmitted from sources to destinations. However, establishing direct connectivity between sources and destinations is not always feasible. To enhance connectivity, a relay, known as an additional communication node, is introduced to facilitate data forwarding. In various applications, such as sensor networks, relay nodes are commonly employed to assist sensors in conserving their limited power while delivering sampled data to destination nodes \cite{Hou_TWC2005, Lloyd_TCOMP2007}. Additionally, with the rise of Industrial IoT networks in the era of Industry 4.0, wireless sensors and actuators are increasingly deployed \cite{Sisinni_TII2018}. In such industrial IoT networks. Having a relay between sensors and controllers becomes crucial in such industrial IoT networks, especially for safety-critical missions that require up-to-date information \cite{Zakeri_Arxiv2022}. Therefore, this study focuses on a scenario where a single relay terminal enables the transmission of fresh information from multiple sensors to their respective destinations in a timely manner. Moreover, we consider that the communication links between the sensors and the relay, as well as between the relay and the destinations, are prone to errors. Hence, transmission errors must be taken into account when optimizing the AoI at the destination. Furthermore, since a relay is shared by multiple sensor-destination pairs over a wireless channel, there is a limit to the number of concurrent communication links. In other words, careful scheduling of actions for sensors and the relay is necessary to maximize the efficiency of the single relay terminal. Consequently, the relay needs to make decisions regarding which sensors should transmit new status information and which destination nodes the relay should update. As destination nodes are updated via the relay, it is crucial for the relay to have access to fresh information. Therefore, it is natural to have the relay jointly select which sensors to sample and which destination nodes to update in a coordinated manner. In this context, we refer to the relay terminal as a coordinating relay, responsible for scheduling the sampling and updating of new statuses.
	
	In this regard, the objective of our paper is to determine the optimal policy that includes sampling and updating decisions to minimize the AoI at the destination nodes. Specifically, we aim to minimize the AoI at the destination nodes when a limited number of simultaneous transmissions are allowed. To account for the varying importance of physical processes, we assign weights to each pair of sensor and destination node that samples and monitors a specific physical process. As a special case, when all pairs are equally important and transmission errors do not occur, we characterize the necessary and sufficient conditions for optimality. By utilizing this condition, we derive a closed-form expression for the fundamental limit of the average sum AoI. We find that the minimum average sum AoI does not increase over time, regardless of the number of allowed concurrent transmissions, and that an optimal policy that achieves the minimum average sum AoI is a greedy policy. Moreover, in error-prone networks with identical importance assigned to every pair, we also prove that our proposed greedy policy is optimal. Furthermore, for IoT networks where the importance of sensors varies, we propose a policy obtained from deep reinforcement learning. Our contributions can be summarized as follows:
		\begin{itemize}
			\item We formulate a joint optimization problem for multiple sensors-destination pairs and a single relay, aiming to minimize the average weighted sum AoI at the destination nodes. This involves optimizing both the sampling and updating policy.
			\item  In an error-free symmetric network where the AoI at different sensors carries equal importance (equal weights), we derive necessary and sufficient conditions for the optimal policy.
			\item We propose a greedy policy that satisfies the optimal conditions.
			\item We provide a closed-form characterization of the average sum AoI with the greedy policy. We demonstrate that, regardless of the allowed number of transmissions in each time, the average sum AoI followed by the greedy policy does not increase after a given time.
			\item In an error-prone symmetric network, we show that the optimal policy samples the sensors with the largest AoI up to the sampling capability, and updates the destination nodes with the largest AoI gap up to the updating capability.
			\item We propose a deep Q-network (DQN) policy utilizing deep reinforcement learning to achieve a low average weighted sum AoI at the destination nodes in a network with arbitrary weights and error probabilities.
	\end{itemize}
	
	\subsection{Related Work} 
	Assuming that new status are given randomly,  \cite{Kaul_INFOCOM2012,Yates_TIT2019,Hsu_TMC2019, Kalør_WCL2019, Javani_Globecom2019, Kosta_JCN2019, Sun_TCOM2020, Bastopcu_TWC2021 } characterized AoI in different types of channels: non-shared independent channels \cite{Kaul_INFOCOM2012, Yates_TIT2019}, broadcasting channels \cite{Hsu_TMC2019, Kalør_WCL2019}, and random access channel \cite{Javani_Globecom2019, Kosta_JCN2019, Sun_TCOM2020}. For edge caching system, when a user updates a content via a cache, \cite{Bastopcu_TWC2021}  investigated the optimal average rate of update requests at the cache and the user to minimize the average freshness of files when a user updates content through a cache. In another line of research \cite{Zhou_TCOM2019, Abd_TCOM2020, Ceran_INFOCOM2019, Kadota_TNET2018, Chen_Arxiv2019, Chen_Arxiv2020},  the deterministic generation of new status updates is considered in the generate-at-will model. Transmission protocols dependent on the value of AoI were proposed in \cite{Chen_Arxiv2019, Chen_Arxiv2020}. where nodes send packets with a probability. Additionally,  in \cite{Ceran_INFOCOM2019, Abd_TCOM2020}, the authors  studied power-efficient status update policies for IoT sensors with limited power and energy-harvesting capabilities. For broadcast channels, \cite{Kadota_TNET2018} analyzed and compared the performance of different scheduling policies assuming AoI increases by units of frames defined as a given time duration. The cost for obtaining fresh information to reduce AoI was taken into consideration in \cite{Zhou_TCOM2019}.
	
	Recently, some initial studies considered AoI in relay networks \cite{Simiscuka_IWCMC2018, Cao_ICC2020, Zhou_WCSP2019, Hu_IOTJ2020, Talak_Allerton2017, Moradian_WCNC2020, Maatouk_ITW2018, Arafa_TWC2019, Sun_INfocom2018, Beytur_BLACK}. In \cite{Simiscuka_IWCMC2018}, AoI is used as one of the evaluation criteria for designing IoT multimedia systems with a relay. AoI-minimizing system design was investigated in \cite{Cao_ICC2020, Zhou_WCSP2019, Hu_IOTJ2020} when unmanned-aerial vehicles (UAVs) take the role of relay between sensors and destination nodes. For a single-pair of a sensor and a destination node, \cite{Cao_ICC2020} formulated an optimization problem to jointly determine the trajectory and communication strategy for minimizing the peak AoI in a single-pair scenario. \cite{Zhou_WCSP2019} focused on trajectory optimization to minimize the sum AoI for multiple sensors and a single destination. \cite{Hu_IOTJ2020} assumed UAVs can transfer energy to sensors, and a joint strategy for positioning, energy transfer, and data collection was investigated. In \cite{Talak_Allerton2017}, AoI-minimizing problem was considred in a multi-hop network with interference, leading to the study of a policy that minimizes AoI by determining a subset of pairs that can communicate.  Average AoI in a wireless relay network consisting of a source, a relay, and a destination node is analyzed in \cite{Moradian_WCNC2020, Maatouk_ITW2018, Arafa_TWC2019}. In \cite{Moradian_WCNC2020}, the transmission probability was optimized under a random access protocol.  \cite{Maatouk_ITW2018} characterized the optimal generation rate of updates, while \cite{Arafa_TWC2019} considered an energy-harvesting system where a source and a relay utilize harvested energy to deliver updates to a destination node. \cite{Arafa_TWC2019} investigated an AoI-minimizing transmission policy for the source and the relay, assuming prior knowledge of the time when energy is harvested.  For multiple source and destination pairs, \cite{Sun_INfocom2018}  explored the scheduling of updates when an update is generated randomly. It was shown that selecting the last generated update from the source with the maximum age is optimal in a probabilistic sense when the scheduler can choose a single source to update at each time. \cite{Beytur_BLACK} proposed the maximum-age difference scheduling policy for a server that selects one source to update.
	
	The rest of this paper is organized as follows. In Section \ref{sec:system_model}, we present the sensor network model in consideration. Our objective function and the corresponding optimization problem are presented in Section \ref{sec:problem_formulation}. Section \ref{sec:analysis} provides the mathematical analysis of this problem. Moreover, we verify our analysis via numerical results in Section \ref{sec:simulations}. Finally, Section \ref{sec:conclusion} concludes the paper.
	\section{System Model}\label{sec:system_model}
	\begin{figure}
		\centering
		\includegraphics[scale=0.26]{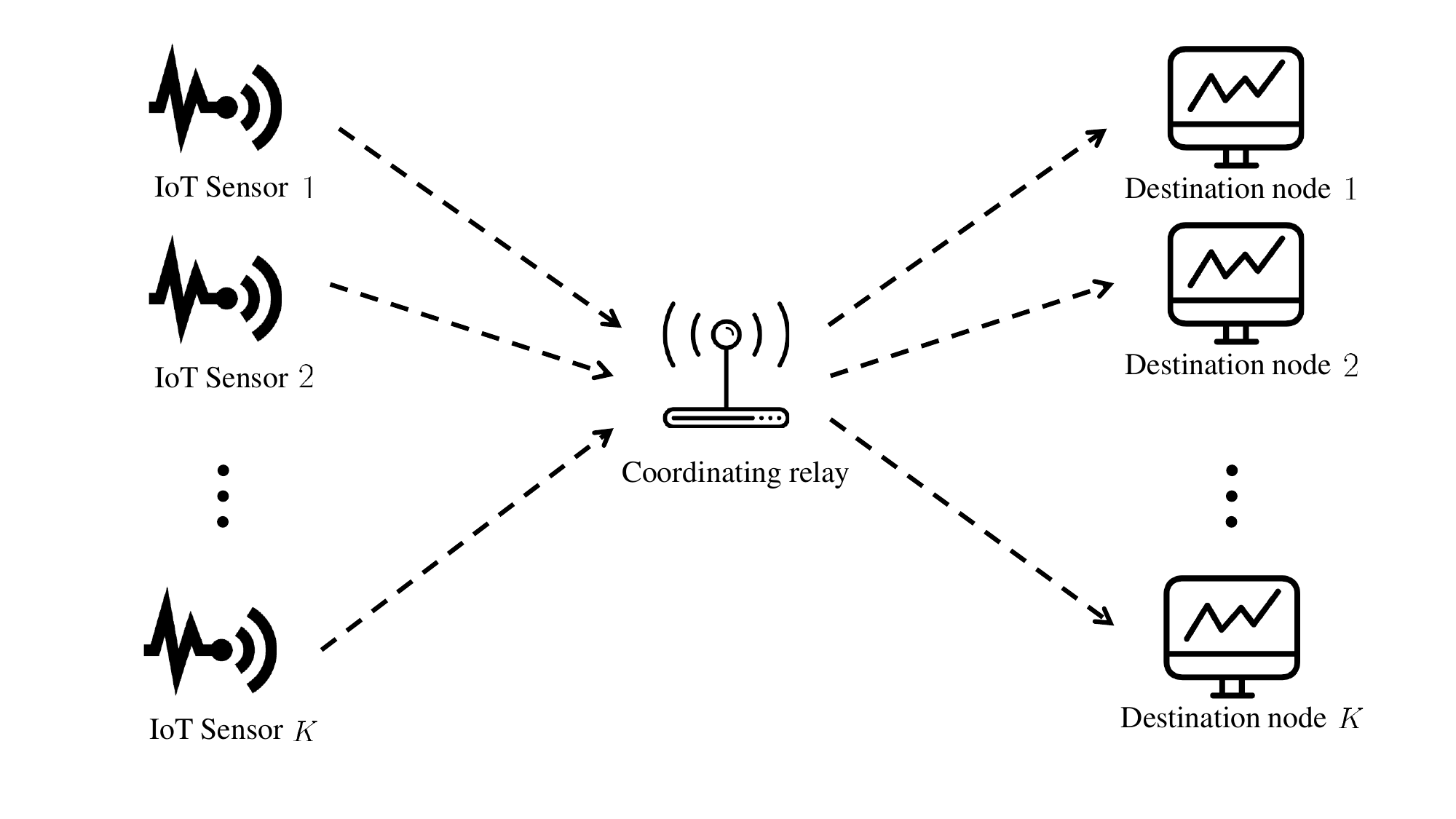}
		\caption{IoT sensor network with a coordinating relay where each sensor observes a distinct physical process which is monitored by a corresponding destination node. The observed information is forwarded from sensors to destination nodes via a coordinating relay}\label{fig:system_model}
	\end{figure}
	We consider a sensor network consisting of $K$ sensors, a relay terminal called as a coordinating relay, and $K$ destination nodes as illustrated in Fig. \ref{fig:system_model}. Each sensor observes a distinct time-varying physical process, and is responsible for conveying the state of this process to a corresponding destination node. As the status of the underlying physical processes change over time, the status information available at the destination nodes become stale, and it is necessary to deliver status updates to the destination nodes. We assume that direct communication links from the sensors to their destinations are not available due to practical limitations, e.g., physical distance, blockage, mismatch in communication technology or hardware, \cite{Hou_TWC2005, Lloyd_TCOMP2007}, and a coordinating relay terminal can forward the status updates to their destination nodes. We assume that the coordinating relay is capable of receiving and transmitting simultaneously\footnote{If we consider half duplex, AoIs at the destination nodes may increase by one time slot due to the half duplex operation.}; however, only a limited number of sensors and destination nodes can communicate with the coordinating relay at each time due to limited wireless resource. Let us define the maximum number of transmissions allowed in the channel from the sensors to the cooperating relay, and from the cooperating relay to the destination nodes as $S$ and $U$, respectively. To avoid the trivial scenarios, we consider the case of $S < K$ and $U < K$. The coordinating relay is equipped with a buffer that can effectively store and update the latest status information for every physical process.\footnote{When the buffer size is enough to store contents, the coordinating relay can work as a edge caching node,  serving and updating contents demanded from destination nodes} Thus, the received updates from the sensors can be stored and delivered in any time after reception. Moreover, the status information stored at the coordinating relay but not delivered to its destination is replaced when a new status update for the same process is received. In this work, we assume that the coordinating relay buffer is large enough to keep track of the status of all the $K$ processes, The coordinating relay buffer limitation will be studied in our future work. Furthermore, we consider the \textit{generate-at-will} model, where the sensors can generate a new status update whenever they are chosen by the coordinating relay \cite{Zhou_TCOM2019, Farazi_ISIT2018}.
	
	Due to the fluctuation of wireless channels over time, not all communications are successful. Consequently, the reception of sampled data from sensors or the delivery of updates to destination nodes can fail due to wireless randomness. In our paper, we primarily focus on communication errors caused by outages, which occur when the wireless channel gain falls below a certain threshold. As a result, the distribution of errors is independent for different communication links, as the wireless channels are independent from each other. Let us define the error probability of the communication link between the coordinating relay and sensor $k$ as $p_k \in [0, 1)$, where $p_k$ represents the likelihood of an error occurring. Similarly, $q_k \in [0, 1)$ represents the error probability of the communication link between the coordinating relay and destination node $k$. It is important to note that both $p_k$ and $q_k$ are constants since the channel fading distribution remains fixed. 
	
	In order to measure timeliness of update, we define the AoI of a process at the coordinating relay, and at the destination node as the  time elapsed  since the corresponding sensor had sampled the status that was delivered most recently to the coordinating relay and to the corresponding destination node, respectively. In this work, a time-slotted model is adopted. Time slots are indexed by positive integer $t \in \mathbb{Z}^+$. If we define  the AoI of process $k$ at the coordinating relay, and at the destination node at  time slot $t$ as $g_k(t)$ and $h_k(t)$, respectively,
	\begin{align}
		g_k(t) = t - \eta_k(t), \label{eq:def_aoi} \\
		h_k(t) = t - \zeta_k(t) , \label{eq:def_aoi_2}
	\end{align} where $\eta_k(t)$ and $\zeta_k(t)$ are  the times when  sensor $k$ generated the most-recent update that  the coordinating relay and the destination node $k$ possess, respectively.
	
	According to \eqref{eq:def_aoi} and \eqref{eq:def_aoi_2}, as time goes on, the information becomes stale and AoI grows.  Thus, appropriately sampling and updating is necessary to keep  information fresh. We assume that communication delay is given as a single time slot; thus, AoI decreases after one time-slot from sampling and updating. Therefore, the status sampled in each time slot cannot be delivered to the destination nodes in the same time slot of sampling. When sensor $k$ samples and transmits to the coordinating relay at time  slot $t-1$,  the AoI at the coordinating relay in time slot $t$ becomes the minimum value of $1$ by definition in \eqref{eq:def_aoi} unless a transmission error occurs. However, if an outage happens, fresh status cannot be transferred to the coordinating relay; hence, the AoI increases by $1$. In other words, if we define $\mathcal{S}(t)$ as a set of sensors to sample at time slot $t$, the evolution of the AoI at the coordinating relay can be represented as, for $k \in \mathcal{S}(t)$,
	\begin{align}
		g_k(t+1) &= \left\lbrace \begin{array}{cl}
			1 & \text{with probability } 1 - p_k \\
			g_k(t) + 1	& \text{with probability } p_k
		\end{array}\right. .\label{eq:evol_ap}
	\end{align}
	For the other processes which are not chosen to be sampled, AoI increases by $1$. For $k \notin \mathcal{S}(t)$,
	\begin{align}
		g_k(t+1) = g_k(t) + 1.
	\end{align}
	
	Regarding the AoI at the destination node, when update from the coordinating relay is successfully delivered to the destination node, the AoI at the destination node decreases. In this case, since the destination node receives update stored at the coordinating relay not directly from the corresponding sensor, AoI at the destination node is not reduced to $1$ but reduced to the value of AoI at the coordinating relay. If $\mathcal{U}(t)$ represents a set of the destination nodes to be updated at time slot $t$, we have the following evolution of the AoI at the destination nodes. For $k \in \mathcal{U}(t)$,
	\begin{align}
		h_k(t+1) &= \left\lbrace \begin{array}{cl}
			g_k(t) + 1, & \text{with probability } 1 - q_k \\
			h_k(t) + 1	& \text{with probability } q_k
		\end{array}\right. ,\label{eq:evol_des}
	\end{align}
	and for $k \notin \mathcal{U}(t)$,
	\begin{align}
		h_k(t+1) = h_k(t) + 1.
	\end{align}
	
	\section{Problem Formulation}\label{sec:problem_formulation}
	
	In this section, we describe how the AoI at the destination nodes are related with the coordinating relay's decision of sampling and updating. In addition to that, in order to keep the destination node informed in a timely manner, we formulate a minimization problem of the average weighted sum AoI at the destination nodes.
	
	Frequently sampling and updating yields lower AoI values at the destination nodes. However, due to limited communication capability, appropriate selection of sensors and destination nodes for sampling and updating is required. Moreover, the selection of sensors and destination nodes by the coordinating relay should be optimized based on the current AoI at the coordinating relay and destination nodes. Depending on the AoI values, the optimal selection of sensors and destinations nodes should be changed. Thus, we consider a scheduling policy which is defined as follows.
	\begin{definition}
		A scheduling policy $\pi$ is defined as a mapping from current AoI at the coordinating relay and destination nodes to the set of sensors which sample new status and the destination nodes which receive updates from the coordinating relay:
		\begin{align}
			\pi \left( \left\lbrace g_k(t) \right\rbrace_{k=1}^K, \left\lbrace h_k(t) \right\rbrace_{k=1}^K \right) = \left( \mathcal{S}^{\pi}(t), \mathcal{U}^{\pi}(t) \right). 
		\end{align} where $ \mathcal{S}^{\pi}(t)$ and $\mathcal{U}^{\pi}(t)$ are the selected sensors and destination nodes by the scheduling policy $\pi$, respectively. 
	\end{definition}
	
	Taking different importance of  physical processes which each destination node monitors,  we consider  the weighted sum AoI at the destination  in order to measure scalar quantity of freshness. Furthermore, since AoI has randomness incurred by outage during delivery, average AoI needs to be considered. Moreover, as the system evolves over time, the current decision of sampling and updating can affect  AoI in the future, thereby time-averaged metric AoI should be defined.  Furthermore, by adjusting weights, the weighted sum AoI can account for  different importance of each AoI. Consequently, as in the literature such as \cite{Kadota_TNET2018,Yi_INFO2020}, we define the average weighted sum AoI as the objective function to minimize;
		\begin{align}
			V^{\pi}(T) = \frac{1}{T}\sum_{t=1}^T \mathbb{E} \left[ \sum_{k=1}^K w_k h^{\pi}_k(t) \right] ,
		\end{align} where $T$ is a finite time-horizon of our interest, $w_k$ is weight for IoT sensor $k$, and $h_k^{\pi}(t)$ is the AoI at destination node $k$ at time slot $t$ with the policy $\pi$.
	
	As we have constraints on the number of sensors and destinations nodes for scheduling in each time, the minimization of the average weighted sum AoI can be formulated as
	\begin{align}
		\nonumber & \min_{\pi} V^{\pi}(T) \\
		\text{s.t. } & \left| \mathcal{S}^{\pi}(t) \right| \leq S, \hspace{5pt} \forall t, \\
		& \left| \mathcal{U}^{\pi}(t) \right| \leq U, \hspace{5pt} \forall t.
	\end{align}
	
	\section{Optimal Policy for Minimizing Average Weighted Sum AoI}\label{sec:analysis}
	In this section, we investigate the optimal policy on sampling and updating to minimize the average weighted sum AoI. Our objective function is the weighted sum AoI at the destination nodes which is directly dependent on the updating policy. When the coordinating relay updates a destination node, the AoI at the destination node is set to be the AoI of the corresponding process at the coordinating relay. In addition to that, the AoI at the coordinating relay is a function of sampling policy. Consequently, sampling and updating policies should be jointly taken into account to reduce the AoI at the destination nodes. Furthermore, each of sampling and updating decision contains a selection of multiple sensors and destination nodes in each time slot. Therefore, all the combinations of sensors and destination nodes should be compared to find the optimal set. As a result, our problem is a joint combinatorial optimization which is known to be hard to solve. To gain an insight on the optimal policy, we first investigate a  simple scenario where the numbers of sensors to sample and destination nodes to update are equal, the importance of different sensors is identical, and the wireless channel gain is high enough to neglect transmission errors.
	
	\subsection{Optimal policy in an error-free symmetric IoT network}
	In this subsection, we consider an error-free symmetric IoT network, where $S = U$, $p_k = q_k = 0$, and $w_k = 1/K$ for any $k$. In this network, as the weights are identically given and the randomness from transmission errors does not exist, the average weighted sum AoI becomes the sum AoI. Thus, we consider the sum AoI as our metric in this subsection. We also set the initial value of the AoI at the coordinating relay and destination nodes to be the same as  $1$ for any physical process (i.e., $g_k(1) = 1$ and $h_k(1) = 1$, $\forall k$). To figure out the optimal policy, we first establish the following lemma.
	\begin{lemma} \label{lem:AoI_ap_vs_AoI_des}
		For any policy $\pi$, $\forall k$ and $\forall t$, the AoI at the destination nodes can be bounded below as
		\begin{align}		
			h_k^{\pi}(t) &\geq g_k^{\pi}(t-1) + 1 , \label{ineq:comp_aoi_ap_des_1} \\
			h_k^{\pi}(t) &\geq g_k^{\pi}(t) . \label{ineq:comp_aoi_ap_des_2}
		\end{align}
	\end{lemma}
	\begin{IEEEproof}	
		Since the evolution of $h_k^{\pi}(t)$ follows \eqref{eq:evol_des}, after at least single successful update, the AoI at the destination node cannot be smaller than the AoI at the coordinating relay. Also, sampling cannot increase AoI at the coordinating relay. Hence, for $t \geq \tau$ such that $k \in \mathcal{U}^{\pi}(\tau)$,
		\begin{align}
			h_k(t) \geq g_k^{\pi}(t-1) + 1.
		\end{align}	
		If $k$ has not been updated at least once until $t$, then $h_k^{\pi}(t) = t$ since  $h_k^{\pi}(1) = 1$. As sampling can only reduce the AoI at the coordinating relay, $g_k^{\pi}(t-1) \leq t-1$. Hence, if $k \notin \bigcup_{\tau=1}^t \mathcal{U}^{\pi}(\tau)$, we can rewrite $h_k^{\pi}(t)$ as
		\begin{align}
			h_k^{\pi}(t) &= t-1 + 1 \\
			& \geq g_k^{\pi}(t-1) + 1
		\end{align}	
		Combining above results, we have, $\forall t$,
		\begin{align}
			h_k^{\pi}(t) \geq g_k^{\pi}(t-1) + 1 .
		\end{align}
		Moreover,  the evolution of AoI at the coordinating relay \eqref{eq:evol_ap} implies that
		\begin{align}
			g_k^{\pi}(t-1) + 1 \geq g_k^{\pi}(t).
		\end{align}
		As a consequence,
		\begin{align}
			h_k^{\pi}(k) \geq g_k^{\pi}(t) .
		\end{align}	
	\end{IEEEproof}
	
	Intuitively, since the AoI at the destination node either increases by $1$ or is set to be equal to the AoI at the coordinating relay, the AoI at the destination node cannot be lower than the AoI at the coordinating relay. Lemma \ref{lem:AoI_ap_vs_AoI_des} confirms this intuition.
	
	Another lemma which will be used to analyze the optimal policy is given below. The following lemma represents the AoI at the coordinating relay and destination nodes as the sum of the AoI reductions by sampling and updating.
	\begin{lemma}\label{lem:sum_AoI}
		The sum AoI at the coordinating relay and at the destination node following a policy $\pi$ can be represented as follows. For any $\pi$ and $t$, 
		\begin{align}
			\sum_{k=1}^K g_k^{\pi}(t) &= tK - \sum_{\tau=1}^{t-1} R(\mathcal{S}^{\pi}(\tau)), \label{eq:sumAoI_ap} \\
			\sum_{k=1}^K h_k^{\pi}(t) &= tK - \sum_{\tau=1}^{t-1} R(\mathcal{U}^{\pi}(\tau)) , \label{eq:sumAoI_des}
		\end{align} where
		\begin{align}
			R(\mathcal{S}^{\pi}(\tau)) &= \sum_{k \in \mathcal{S}^{\pi}(\tau)} g_k^{\pi}(\tau), \\
			R(\mathcal{U}^{\pi}(\tau)) & = \sum_{k \in \mathcal{U}^{\pi}(\tau)} h_k^{\pi}(\tau) - g_k^{\pi}(\tau).
		\end{align}
	\end{lemma}
	\begin{IEEEproof}	
		We prove this lemma using mathematical induction. First, we focus on the proof for the expression of the sum AoI at the coordinating relay. Suppose \eqref{eq:sumAoI_ap} holds for $t>1$. Since the evolution of AoI at the coordinating relay follows \eqref{eq:evol_ap},
		\begin{align}
			\sum_{k=1}^K g_k^{\pi}(t+1) &= \sum_{k \in \mathcal{S}^{\pi}(t)} g_k^{\pi}(t+1) + \sum_{k \notin \mathcal{S}^{\pi}(t)} g_k^{\pi}(t+1) , \\
			& = \sum_{k=1}^K 1 + \sum_{k \notin \mathcal{S}^{\pi}(t)} g_k^{\pi}(t) , \\
			& = K + \sum_{k=1}^K g_k^{\pi}(t) - \sum_{k \in \mathcal{S}^{\pi}(t)} g_k^{\pi}(t). \label{lem2_temp1}
		\end{align}
		By supposition, \eqref{lem2_temp1} can be rewritten as 
		\begin{align}
			\sum_{k=1}^K g_k(t+1) &= K + tK - \sum_{\tau=1}^{t-1} \sum_{k \in \mathcal{S}^{\pi}(\tau)} g_k^{\pi}(\tau) - \sum_{k \in \mathcal{S}(t)} g_k^{\pi}(t), \\
			& = (t+1) K - \sum_{\tau=1}^t \sum_{k \in \mathcal{S}^{\pi}(\tau)} g_k^{\pi}(\tau)
		\end{align}
		Thus, \eqref{eq:sumAoI_ap} holds for $t+1$ if it is valid for $t$.	\\
		Moreover, since $g_k^{\pi}(1) = 1$ for any policy $\pi$, the sum of the AoI at the coordinating relay in $t=1$ becomes
		\begin{align}
			\sum_{k=1}^K g_k^{\pi}(1) = K,
		\end{align} which is identical to \eqref{eq:sumAoI_ap} when $t=1$.
		Therefore, by mathematical induction, \eqref{eq:sumAoI_ap} holds for any $t$.
		Using mathematical induction with similar derivation, we can also prove the expression of the sum AoI at the destination nodes, \eqref{eq:sumAoI_des}, holds.
	\end{IEEEproof}
	
	In fact, $R(\mathcal{S}^{\pi}(t))$ and $R(\mathcal{U}^{\pi}(t))$ are the reductions of AoI by sampling and updating at time slot $t$, respectively. Moreover, $tK$ is the sum AoI when sensors and the destination nodes do not sample or receive any new updates. Thus, Lemma \ref{lem:sum_AoI} implies that sum AoI  at time slot $t$ is the subtraction of the  AoI reductions accumulated until time slot $t-1$  from the sum AoI without  sampling or updating. Based on Lemmas \ref{lem:AoI_ap_vs_AoI_des} and \ref{lem:sum_AoI}, we can derive the following proposition which is essential to have the optimality condition.
	
	\begin{proposition}\label{prop:AoI_reduction}
		For any policy $\pi$ and $\forall t$, the accumulated AoI reductions at the coordinating relay until time  slot $t-1$ are always greater than or equal to the accumulated AoI reductions at the destination nodes until time $t$.
		\begin{align}
			\sum_{\tau=1}^{t-1} R(\mathcal{S}^{\pi}(\tau)) \geq \sum_{\tau=1}^t R(\mathcal{U}^{\pi}(\tau)). \label{ineq:AoI_reduction}
		\end{align}
	\end{proposition}
	\begin{IEEEproof}
		By Lemma \ref{lem:sum_AoI}, given policy $\pi$, the difference of the sum AoI at the destination nodes in the $t+1$-th time slot and at the coordinating relay in the  $t$-th time slot can be represented as
		\begin{align}
			\sum_{k=1}^K \left[ h_k^{\pi}(t+1) - g_k^{\pi}(t) \right] = K + \sum_{\tau=1}^{t-1} R(\mathcal{S}^{\pi}(\tau)) - \sum_{\tau=1}^{t} R(\mathcal{U}^{\pi}(\tau)) . \label{prop1_temp1}
		\end{align}
		Also, \eqref{prop1_temp1} can be rewritten as
		\begin{align}
			\sum_{k=1}^K \left[ h_k^{\pi}(t+1) - \left( g_k^{\pi}(t) + 1 \right) \right] = \sum_{\tau=1}^{t-1} R(\mathcal{S}^{\pi}(\tau)) - \sum_{\tau=1}^{t} R(\mathcal{U}^{\pi}(\tau)). \label{prop1_temp4}
		\end{align}
		In addition to that, Lemma \ref{lem:AoI_ap_vs_AoI_des} implies that
		\begin{align}
			\sum_{k=1}^K \left[h_k(t+1) - \left( g_k(t) +1 \right)\right] \geq 0. \label{prop1_temp2}
		\end{align}
		Combining \eqref{prop1_temp4} and \eqref{prop1_temp2}, we have
		\begin{align}
			\sum_{\tau=1}^{t-1} R(\mathcal{S}^{\pi}(\tau)) \geq \sum_{\tau=1}^{t} R(\mathcal{U}^{\pi}(\tau)) .
		\end{align}
	\end{IEEEproof}
	
	Proposition \ref{prop:AoI_reduction} implicates that, given a policy, the accumulated AoI reductions by updating are bounded above by the accumulated AoI reductions by sampling. Fundamentally, updating reduces the AoI gap between the coordinating relay and the destination nodes by lowering the AoI at the destination nodes, whereas sampling increases the AoI gap between the coordinating relay and the destination nodes by reducing the AoI at the coordinating relay. However, the AoI at the destination nodes cannot be less than the AoI at the coordinating relay. Thus, updating cannot achieve a larger gain than sampling in terms of accumulated AoI reductions. More precisely, since sampling and updating occur at the beginning of each time slot, the AoI gap produced by sampling  in the current time slot cannot be reduced by updating at the same time slot. Consequently, there exists a time gap of at least a single-time slot between sampling and updating. Accounting for the time gap, the accumulated AoI reductions by updating up to time slot $t$ cannot be larger than the accumulated AoI reductions by sampling up to time slot $t-1$.
	
	\begin{algorithm}[t!]
		\caption{Greedy Policy}
		\label{alg:proposed}
		\begin{algorithmic}
			\Require $g(1) =\left[ 1, \cdots, 1\right] $, $h(1) = \left[ 1, \cdots, 1\right]$, $\left\lbrace w_k \right\rbrace_{k=1}^K$
			\For {$t=1, \cdots, T$}
			\State $\mathcal{S}^{G}(t) = \argmax_{\mathcal{S} \subset \mathcal{K}, |\mathcal{S}(t)| \leq S}\sum_{k \in \mathcal{S}} w_k g_k^{G}(t)$
			\State $\mathcal{U}^{G}(t) = \argmax_{\mathcal{U} \subset \mathcal{K}, |\mathcal{U}| \leq U}\sum_{k \in \mathcal{U}} w_k \left( h_k^{G}(t) - g_k^{G}(t) \right)$
			\EndFor
		\end{algorithmic}	
	\end{algorithm}	
	
	\begin{table}
		\caption{Evolution of AoI at the coordinating relay and destination nodes with greedy policy in error-free symmetric IoT network where $K=5$, $S=3$, $U=3$.} \label{tbl:greedy_evol}	
		\begin{center}
			\begin{tabular}{||c| c| c| c ||} 
				\hline
				&$T=1$ & $T=2$ & $T=3$ \\  
				\hline\hline
				$\mathbf{g}(t)$ &$\left[ 1,1,1,1,1 \right]$ & $\left[ 1,1,1,2,2 \right]$ & $\left[ 2,2,1,1,1 \right]$  \\ 
				\hline
				$\mathbf{h}(t)$ &$\left[ 1,1,1,1,1 \right]$ & $\left[ 2,2,2,2,2 \right]$ & $\left[ 2,2,2,3,3 \right]$  \\
				\hline 
				&$T=4$ & $T=5$ & $T=6$ \\ 
				\hline \hline
				$\mathbf{g}(t)$ &$\left[ 1,1,1,2,2 \right]$ & $\left[ 2,2,1,1,1 \right]$ & $\left[ 1,1,1,2,2 \right]$  \\ 
				\hline
				$\mathbf{h}(t)$ &$\left[ 3,3,2,2,2 \right]$ & $\left[ 2,2,2,3,3 \right]$ & $\left[ 3,3,2,2,2 \right]$  \\		
				\hline
			\end{tabular}
		\end{center}	
	\end{table}
	
	Now, we propose a greedy policy and prove that the proposed greedy policy is optimal. The greedy policy is a sampling and updating policy selecting sensors which have the $S$-largest AoIs at the coordinating relay and the destination nodes which have the $U$-largest AoI gaps. The detail of the greedy policy is described in Algorithm \ref{alg:proposed} for  a general scenario where different weights of processes and outages in transmission are considered. Also, the example of AoI evolution with the  greedy policy is demonstrated in Table \ref{tbl:greedy_evol} for error-free symmetric IoT network where $K=5$, $S=3$, $U=3$.
	
	In the following propositions, we figure out important properties of the greedy policy.
	\begin{proposition}\label{prop:greedy_opt_sampling}
		For error-free symmetric IoT network where $S=U$, $p_k = 0, q_k = 0,$ and $w_k = \frac{1}{K}$ for any $k$, the sum AoI at the coordinating relay is minimized by the greedy sampling policy;
		\begin{align}
			G = \arg\min_{\pi} \frac{1}{T K} \sum_{t=1}^T \sum_{k=1}^K g_k^{\pi}(t).
		\end{align} Equivalently,
		\begin{align}
			G = \argmax_{\pi} \sum_{t=1}^T \sum_{\tau=1}^{t-1} R(\mathcal{S}^{\pi}(\tau)).
		\end{align}
	\end{proposition}
	\begin{IEEEproof}	
		From \eqref{eq:evol_ap}, the AoI of  process $k$ at the coordinating relay with a policy $\pi$ evolves
		\begin{align}
			g^{\pi}_k(t) = \mathbbm{1}(k \notin \mathcal{S}^{\pi}(t)) g_k^{\pi}(t-1) + 1,
		\end{align} where $\mathbbm{1}(x)$ is an indicator function which is $1$ if $x$ is true, and $0$, otherwise.
		After some manipulations, we can have general expression of $g_k(t)$ as a function of $g_k(1)$.
		\begin{align}
			g_k^{\pi} (t) = \prod_{\tau=1}^{t-1} \mathbbm{1}(k \notin \mathcal{S}^{\pi}(\tau))g_k^{\pi}(1) + \sum_{\tau=2}^{t-1} \prod_{d = \tau}^{t-1} \mathbbm{1}(k \notin \mathcal{S}^{\pi}(d)) + 1.
		\end{align}
		Accordingly, the sum AoI at the coordinating relay in $t$ can be written as
		\begin{align}
			\sum_{k=1}^K g_k^{\pi}(t) = \sum_{k=1}^K \prod_{\tau=1}^{t-1} \mathbbm{1}(k \notin \mathcal{S}^{\pi}(\tau)) + \sum_{k=1}^K \sum_{\tau=2}^{t-1} \prod_{d = \tau}^{t-1} \mathbbm{1}(k \notin \mathcal{S}^{\pi}(d)) + K, \label{eq:sumAoI_ap_2}
		\end{align} where we use the initial value, $g_k^{\pi}(1) = 1$ for any $\pi$ and $k$.
		Using the definition of the indicator function, the product of indicator functions becomes
		\begin{align}
			\prod_{\tau=1}^{t-1} \mathbbm{1}(k \notin \mathcal{S}^{\pi}(\tau)) = \left\lbrace \begin{array}{ll}
				1 & \text{if } k \notin \bigcup_{\tau=1}^{t-1} \mathcal{S}^{\pi}(\tau) \\
				0 & \text{otherwise}
			\end{array}. \right.
		\end{align}
		Hence, depending on $\bigcup_{\tau=1}^{t-1} \mathcal{S}^{\pi}(\tau)$, the sum AoI at the coordinating relay becomes different. If $\bigcup_{\tau=1}^{t-1} \mathcal{S}^{\pi}(\tau)$ includes all the $K$ sensors, $\sum_{k=1}^K\prod_{\tau=1}^{t-1} \mathbbm{1}(k \notin \mathcal{S}^{\pi}(\tau))$ becomes $0$. Since only $S$ sensors are chosen in each time slot, there exists a limit on the size of $\bigcup_{\tau=1}^{t-1} \mathcal{S}^{\pi}(\tau)$. Thus, we have the lower bound of $\sum_{k=1}^K\prod_{\tau=1}^{t-1} \mathbbm{1}(k \notin \mathcal{S}^{\pi}(\tau))$ as 
		\begin{align}
			\sum_{k=1}^K\prod_{\tau=1}^{t-1} \mathbbm{1}(k \notin \mathcal{S}^{\pi}(\tau)) \geq \min\left\lbrace K - \left(t-1\right) S, 0 \right\rbrace.
		\end{align}
		Also, for the second term in \eqref{eq:sumAoI_ap_2}, we have
		\begin{align}
			\prod_{d = \tau}^{t-1} \mathbbm{1}(k \notin \mathcal{S}^{\pi}(d)) = \left\lbrace \begin{array}{ll}
				1 & \text{if } k \notin \bigcup_{\tau=1}^{t-1} \mathcal{S}^{\pi}(\tau) \\
				0 & \text{otherwise}
			\end{array}. \right.
		\end{align}
		Similarly,
		\begin{align}
			\sum_{k=1}^K\prod_{\tau=1}^{t-1} \mathbbm{1}(k \notin \mathcal{S}^{\pi}(\tau)) \geq \min\left\lbrace K - \left(t-\tau\right) S, 0 \right\rbrace.
		\end{align}
		Therefore, we have the following lower bound on the  sum AoI at the coordinating relay.
		\begin{align}
			\nonumber \sum_{k=1}^K g_k^{\pi}(t) &\geq \min\left\lbrace K - \left(t-1\right) S, 0 \right\rbrace \\ &+ \sum_{\tau=2}^{t-1} \min\left\lbrace K - \left(t-\tau\right) S, 0 \right\rbrace + K.
		\end{align}
		As the greedy sampling policy chooses sensors in turn, the  cardinality of $\bigcup_{\tau=1}^{t-1} \mathcal{S}^{G} (\tau) $  achieved by the greed sampling policy equals to $K - \left(t-1\right)S$ for any $t$. Hence, the greedy sampling policy satisfies
		\begin{align}
			\sum_{k=1}^K\prod_{\tau=1}^{t-1} \mathbbm{1}(k \notin \mathcal{S}^{G}(\tau)) &= \min\left\lbrace K - \left(t-1\right) S, 0 \right\rbrace ,\\ \sum_{k=1}^K\prod_{d=\tau}^{t-1} \mathbbm{1}(k \notin \mathcal{S}^{G}(d)) &= \min\left\lbrace K - \left(t-\tau\right) S, 0 \right\rbrace .  
		\end{align}
		Therefore, the greedy policy obtains the minimum sum AoI at the coordinating relay for any $t$.
	\end{IEEEproof}
	
	\begin{proposition}\label{prop:greedy_reduction}
		For error-free symmetric IoT network where $S=U$, $p_k = 0, q_k = 0,$ and $w_k = \frac{1}{K}$ for any $k$, the AoI reductions by sampling and updating via the greedy policy are given, respectively, as 
		\begin{align}
			R(\mathcal{S}^{G}(t)) &= \min\left\lbrace tS, K\right\rbrace , \\
			R(\mathcal{U}^G(t)) &= \min\left\lbrace (t-1)U, K\right\rbrace .
		\end{align}
		Furthermore, the greedy policy achives that the accumulated AoI reduction at the coordinating relay until time slot $t-1$ is equal to the accumulated AoI reduction at the destination nodes until time $t$. In other words, the greedy policy satisfies \eqref{ineq:AoI_reduction} in Proposition \ref{prop:AoI_reduction} with equality.
		\begin{align}
			\sum_{\tau=1}^{t-1} R(\mathcal{S}^{G}(\tau)) = \sum_{\tau=1}^{t} R(\mathcal{U}^{G}(\tau)). \label{eq:greedy_achieve}
		\end{align} 
	\end{proposition}
	\begin{IEEEproof}	
		First, we consider AoI reduction at the coordinating relay by the greedy sampling. For $t \leq \frac{K}{S}$, there exists $K - \left(t-1\right) S$ processes which have never been sampled. Thus, those processes have the largest AoI at the coordinating relay.
		In addition, for given initial AoI $g_k(1) = 1$ for $\forall k$, $g_k(t) = t$ for $k \notin \bigcup_{\tau=1}^{t-1} \mathcal{S}^G(\tau)$ and $t \leq \frac{K}{S}$. If there exists more than $S$ processes of which AoI is $t$, the greedy policy samples $S$ processes with AoI $t$. In other words, if $tS \leq K$,
		\begin{align}
			R(\mathcal{S}^{G}(t)) = tS.
		\end{align}
		On the other hand, if $(t-1) S \leq K < t S$, some of the sensors are chosen the first time and others were chosen once. In this case, the sensors which were sampled at $t=1$ have the second largest AoI of the corresponding process since it has been the longest time from the first sampling time. Moreover, the AoI of the processes sampled at $t=1$ is $t-1$. Thus, the AoI reduction by sampling is given as
		\begin{align}
			R(\mathcal{S}^{G}(t)) &= \left( K - (t-1)S \right) t + \left( S - \left( K - (t-1)S \right) \right) (t-1) , \\
			& = K.
		\end{align}
		Therefore, for $t \leq \frac{K}{S}$, $R(\mathcal{S}^{G}(\tau)) = \min \left\lbrace tS, K\right\rbrace$.\\	
		When $(t-1) S > K $, all the sensors have been chosen to sample at least once. Thus, the maximum-AoI process becomes the one  sampled earliest. In this case, since AoI of processes becomes the difference of current time and the time when it was sampled, the AoIs of processes have one of the value from $1$ to $t'$ where $t' = \left\lceil \frac{K}{S} \right\rceil$. Furthermore, there exist $S$ processes for each value of AoI from $1$ to $t'- 1$. The rest $K - (t'-1) S$ processes have AoI of $t'$. Hence, the greedy sampling chooses $K - (t'-1) S$ sensors with AoI $t'$ and $S - (K - t'S)$ sensors with AoI $t'-1$. Therefore, the reduction of AoI by sampling becomes
		\begin{align}
			R(\mathcal{S}^G(t)) & = t'\left(K - (t'-1)S\right) + (t'-1)\left(S - \left(K - (t'-1)S \right) \right) , \\
			& = K.
		\end{align}
		As $ t \leq \frac{K}{S}$ and $ t > \frac{K}{S}$ are equivalent to $t S \leq K$ and $ tS > K$, the AoI reduction by sampling is given as
		\begin{align}
			R(\mathcal{S}^{G}(t)) &= \min\left\lbrace tS, K\right\rbrace. \label{eq:red_rel}
		\end{align}	
		The AoI reduction by updating can be derived similarly.
		Since updating makes the AoI at the destination nodes equal to AoI at the coordinating relay, the AoI gap becomes zero. Therefore, before sampling, updated processes have zero AoI gap between at the coordinating relay and the destination nodes. In short, the AoI gap is generated by sampling and reduced by updating. Moreover, when $S=U$, the greedy policy selects the same set of processes which were sampled in the previous time to update in the current time because only those set of processes have non-zero AoI gap.\\	
		For $t \leq \frac{K}{U}+1$, since not all the processes are sampled until time slot $t-1$, the AoIs of the processes, which were sampled in time slot $t-1$ so that their AoIs were refreshed to 1 at the coordinating relay, is $t$  at the destination node. Thus, the AoI gap becomes $t-1$. As a result, we have the following AoI reduction by updating.
		\begin{align}
			R(\mathcal{U}^{G}(t)) = (t-1)U.
		\end{align}	
		For $t > \frac{K}{U}+1$, as explained before, $K - (t'-1) S$ processes with AoI $t'$ at the coordinating relay and $S - (K - (t'-1)S)$ processes with AoI $t'-1$ at the coordinating relay were sampled in $t-1$ and their AoIs at the destination nodes are the same as the AoI at the coordinating relay. Therefore, in time slot $t$, $K - (t'-1) S$ processes have the AoI gap of $t'$ and $S - (K - (t'-1)S)$ processes have the AoI gap of $t'-1$. Thus, the AoI reduction by updating can be represented as
		\begin{align}
			R(\mathcal{U}^{G}(t)) &= t'(K-t'S) + (t'-1)(S - (K-t'S)) , \\
			& = K.
		\end{align}
		As $ t \leq \frac{K}{U} + 1$ and $ t > \frac{K}{U} + 1$ are equivalent to $(t-1) U \leq K$ and $ (t-1)U > K$, the AoI reduction by updating becomes
		\begin{align}
			R\left(\mathcal{U}^{G}(t)\right) &= \min\left\lbrace (t-1)U, K\right\rbrace. \label{eq:red_des}
		\end{align}
		Now, we prove that the greedy policy satisfies \eqref{ineq:AoI_reduction} with equality. From \eqref{eq:red_rel} and \eqref{eq:red_des}, as $S=U$, we have for $t>1$,
		\begin{align}
			R(\mathcal{S}^{G}(t)) = R(\mathcal{U}^{G}(t+1)).
		\end{align} Thus, the accumulated AoI reductions at the destination nodes can be expressed as
			\begin{align}
				\sum_{\tau=1}^{t-1} R(\mathcal{S}^{G}(\tau)) &= \sum_{\tau=1}^{t-1} R(\mathcal{U}^{G}(\tau+1) \\
				&= \sum_{\tau=2}^{t} R(\mathcal{U}^{G}(\tau)). \label{eq:temp1}
			\end{align}
			Since we assume that initial value of AoI is set to be $1$. we have $g_k(0) = h_k(0) = 1$. This  implies that $R(\mathcal{U}^{G}(1)) = 0$. Therefore, we have
			\begin{align}
				\sum_{\tau=1}^{t-1} R(\mathcal{S}^{G}(\tau)) &=\sum_{\tau=2}^{t} R(\mathcal{U}^{G}(\tau) + R(\mathcal{U}^{G}(1)), \\
				& = \sum_{\tau=1}^{t} R(\mathcal{U}^{G}(\tau)).
		\end{align}
	\end{IEEEproof}
	
	In the following theorem, we figure out the necessary and sufficient condition for the optimal policy.
	
	\begin{theorem}[Necessary and sufficient condition of optimal policy] \label{thm:opt_condition} For error-free symmetric IoT network, a policy $\pi^*$ is optimal if and only if $\pi^*$ satisfies,
		\begin{align}
			\sum_{t=1}^{T} \sum_{\tau = 1}^{t-2} R(\mathcal{S}^{\pi^*}(t)) &= \max_{\pi} \sum_{t=1}^{T} \sum_{\tau = 1}^{t-2} R(\mathcal{S}(t)) , \label{eq:optimality_1} \\
			\sum_{\tau=1}^{t-1} R(\mathcal{S}^{\pi^*}(\tau)) &= \sum_{\tau=1}^t  R(\mathcal{U}^{\pi^*}(\tau)), \hspace{10 pt}  \forall t. \label{eq:optimality_2}
		\end{align}
	\end{theorem}
	\begin{IEEEproof}	
		By Lemma \ref{lem:sum_AoI}, the sum AoI at the destination nodes for any policy $\pi$ can be expressed as
		\begin{align}
			\frac{1}{T K} \sum_{t=1}^T \sum_{k=1}^K h^{\pi}_k(t) = \frac{1}{T K} \sum_{t=1}^T\left[ tK - \sum_{\tau=1}^{t-1} R(\mathcal{U}^{\pi}(\tau)) \right] . \label{thm_temp1}
		\end{align}
		Also, due to Proposition \ref{prop:AoI_reduction}, \eqref{thm_temp1} can be bounded below as
		\begin{align}
			\frac{1}{T K} \sum_{t=1}^T \sum_{k=1}^K h^{\pi}_k(t) \geq \frac{1}{TK} \sum_{t=1}^T \left[ tK - \sum_{\tau=1}^{t-2} R(\mathcal{S}^{\pi}(\tau)) \right].
		\end{align}
		The lower bound is minimized when $\sum_{\tau=1}^{t-2} R(\mathcal{S}^{\pi}(\tau))$ is maximized. As a consequence, for any policy $\pi$, the sum AoI cannot be lower than the minimum of lower bound.
		\begin{align}
			\frac{1}{T K } \sum_{t=1}^T \sum_{k=1}^K h^{\pi}_k(t) \geq \min_{\pi'} \frac{1}{T K} \sum_{t=1}^T \left[ tK - \sum_{\tau=1}^{t-2} R(\mathcal{S}^{\pi'}(\tau)) \right]. \label{eq:thm_temp1}
		\end{align}
		Furthermore, as the first term of right-hand side (RHS) of \eqref{eq:thm_temp1} is irrelevant with $\pi'$, we can rewrite RHS of \eqref{eq:thm_temp1} as
		\begin{align}
			\nonumber &\frac{1}{TK} \min_{\pi'} \sum_{t=1}^T \left[ tK - \sum_{\tau=1}^{t-2} R(\mathcal{S}^{\pi'}(\tau)) \right] \\
			& = \frac{1}{TK} \left( \sum_{t=1}^T tK - \max_{\pi'} \sum_{t=1}^T \sum_{\tau=1}^{t-2} R(\mathcal{S}^{\pi'}(\tau)) \right).
		\end{align}
		If there exists a policy $\pi^*$ which satisfies \eqref{eq:optimality_1} and \eqref{eq:optimality_2} simultaneously can have the minimum sum AoI.
		For any $\pi$, 
		\begin{align}
			\frac{1}{T K}\sum_{t=1}^T \sum_{k=1}^K h^{\pi}_k(t) &= \frac{1}{T K} \sum_{t=1}^T \left[ tK - \sum_{\tau=1}^{t-2} R(\mathcal{U}^{\pi}(\tau)) \right], \\
			& \geq \frac{1}{T K} \left( \sum_{t=1}^T tK - \sum_{t=1}^T \sum_{\tau=1}^{t-2}   R(\mathcal{S}^{\pi}(\tau)) \right), \\
			& \geq \frac{1}{T K} \left( \sum_{t=1}^T tK - \max_{\pi'} \sum_{t=1}^T \sum_{\tau=1}^{t-2}   R(\mathcal{S}^{\pi'}(\tau)) \right), \\
			& = \frac{1}{T K}\sum_{t=1}^T \sum_{k=1}^K h^{\pi^*}_k(t)
		\end{align} 
		Therefore, a policy which achieves \eqref{eq:optimality_1} and \eqref{eq:optimality_2} is optimal.\\	
		On the other hand, to prove the converse, suppose there exists an optimal policy $\pi^*$ which does not satisfy \eqref{eq:optimality_1} or \eqref{eq:optimality_2}. Then, 
		\begin{align}
			\sum_{t=1}^T \sum_{\tau=1}^{t-2} R(\mathcal{S}^{\pi^*}(\tau)) < \max_{\pi} \sum_{t=1}^T \sum_{\tau=1}^{t-2} R(\mathcal{S}^{\pi}(\tau)), \label{ineq:thm_temp1}
		\end{align} or, by  Proposition \ref{prop:AoI_reduction},
		\begin{align}
			\sum_{\tau=1}^{t-1} R(\mathcal{S}^{\pi^*}(\tau)) > \sum_{\tau=1}^t  R(\mathcal{U}^{\pi^*}(\tau)). \label{ineq:thm_temp2}
		\end{align}	
		However, as we showed in Propositions \ref{prop:greedy_opt_sampling} and \ref{prop:greedy_reduction}, there already exists a greedy policy $G$ which satisfies both \eqref{eq:optimality_1} and \eqref{eq:optimality_2}. 
		Hence, the sum AoI for policy $\pi^*$ can be represented as
		\begin{align}
			\frac{1}{T K}\sum_{t=1}^T \sum_{k=1}^K h^{\pi^*}_k(t) &=  \frac{1}{T K} \sum_{t=1}^T\left[ tK - \sum_{\tau=1}^{t-1} R(\mathcal{U}^{\pi^*}(\tau)) \right] , \\
			& \geq  \frac{1}{T K} \left( \sum_{t=1}^T tK -  \sum_{t=1}^T\sum_{\tau=1}^{t-2} R(\mathcal{S}^{\pi^*}(\tau)) \right)  , \label{ineq:thm_temp3} \\
			& \geq \frac{1}{T K} \left( \sum_{t=1}^T tK -  \max_{\pi'} \sum_{t=1}^T \sum_{\tau=1}^{t-2}  R(\mathcal{S}^{\pi'}(\tau)) \right) , \label{ineq:thm_temp4} \\
			& = \frac{1}{T K} \left( \sum_{t=1}^T tK -  \sum_{t=1}^T\sum_{\tau=1}^{t-2} R(\mathcal{S}^{G}(\tau)) \right)  , \\
			& = \frac{1}{T K} \sum_{t=1}^T \sum_{k=1}^K h^{G}_k(t).
		\end{align}
		If either \eqref{ineq:thm_temp1} or \eqref{ineq:thm_temp2} holds, one of the inequalities, \eqref{ineq:thm_temp3} or \eqref{ineq:thm_temp4} should hold without equality. Consequently, we have
		\begin{align}
			\frac{1}{T K}\sum_{t=1}^T \sum_{k=1}^K h^{\pi^*}_k(t) > \frac{1}{T K} \sum_{t=1}^T \sum_{k=1}^K h^{G}_k(t). \label{ineq:thm_temp5}
		\end{align}
		However, \eqref{ineq:thm_temp5} contradicts to the optimality of $\pi^*$. By contradiction, any optimal policy should satisfy both \eqref{eq:optimality_1} and \eqref{eq:optimality_2}.
	\end{IEEEproof}
	
	As shown in  Proposition \ref{prop:AoI_reduction}, the accumulated AoI reductions by updating are bounded above by that of sampling. Thus, the AoI reductions at the coordinating relay by a sampling policy which maximizes the AoI reductions at the coordinating relay become the maximum of the upper bound of the AoI reductions at the destination nodes. In this context, if there exists a policy which enables to reduce AoI at the destination by updating as much as the maximized AoI reduction at the coordinating relay by sampling, the policy achieves the minimum sum AoI at the destination nodes, thereby it is optimal. Furthermore, from the Propositions \ref{prop:greedy_opt_sampling} and \ref{prop:greedy_reduction}, the greedy policy is able to achieve the largest AoI reduction at the coordinating relay and the same amount of AoI reduction at the destination nodes. To this end, we can conclude that any other policy which cannot reduce as much as greedy policy does cannot be optimal.
	
	Combining the results of Propositions \ref{prop:greedy_opt_sampling} and \ref{prop:greedy_reduction} with Theorem \ref{thm:greedy_optimality}, we have the optimality of the greedy policy for minimizing the both the average and instantaneous sum AoI at the destination in error-free symmetric IoT network.  
	
	\begin{theorem}[Optimality of Greedy Policy]\label{thm:greedy_optimality}
		For error-free symmetric IoT network where $S=U$, $p_k = 0$, $q_k = 0$, and $w_k = \frac{1}{K}$ for any $k$, the greedy sampling and updating policy described in Algorithm \ref{alg:proposed} is optimal for minimizing average sum AoI at the destination node.
		\begin{align}
			G = \arg\min_{\pi} \frac{1}{T K} \sum_{\tau=1}^T \sum_{k=1}^K h_k^{\pi}(t).
		\end{align}
		Furthermore, the greedy policy also minimizes the instantaneous sum AoI at the coordinating relay and at the destination nodes. 
		\begin{align}
			G &= \arg\min_{\pi} \frac{1}{K} \sum_{k=1}^K h_k^{\pi}(t), \\
			G & = \arg\min_{\pi} \frac{1}{K} \sum_{k=1}^K g_k^{\pi}(t) .
		\end{align}
	\end{theorem}
	\begin{IEEEproof}
		Since we have figured out the necessary and sufficient condition for the  optimal policy in Theorem \ref{thm:opt_condition}, we can prove the optimality of the greedy policy by checking the optimality condition.
		From Proposition \ref{prop:greedy_opt_sampling},
		\begin{align}
			G = \argmax_{\pi} \sum_{t=1}^T \sum_{\tau=1}^{t-1} R(\mathcal{S}^{\pi}(\tau)).
		\end{align}	
		Moreover, by Proposition \ref{prop:greedy_reduction},
		\begin{align}
			\sum_{\tau=1}^{t-1} R(\mathcal{S}^{G}(\tau)) & = \sum_{\tau=1}^t R(\mathcal{U}^{G} (\tau)).
		\end{align}
		As a result, by Theorem \ref{thm:opt_condition}, the greedy policy is optimal for minimizing the average sum AoI in error-free symmetric IoT network.\\	
		Furthermore, from Lemma \ref{lem:sum_AoI} and Proposition \ref{prop:greedy_reduction}, we have
		\begin{align}
			\sum_{k=1}^K h_k^{\pi}(t) &= tK - \sum_{\tau=1}^{t-1} R \left( \mathcal{U}^{\pi}(\tau) \right), \\
			& \geq  tK - \sum_{\tau=1}^{t-2} R \left( \mathcal{S}^{\pi}(\tau) \right) , \\
			& \geq  tK - \max_{\pi} \sum_{\tau=1}^{t-2} R \left( \mathcal{S}^{\pi}(\tau) \right), \\
			& = tK - \sum_{\tau=1}^{t-2} R \left( \mathcal{S}^{G}(\tau) \right).
		\end{align}
		Thus, the greedy policy also achieves the minimum instantaneous sum AoI at the destination nodes.
	\end{IEEEproof}
	
	Since the AoI reductions by sampling and updating are maximized at any time with the greedy policy, we can obtain the minimum sum AoI at the coordinating relay and destination nodes.
	\begin{theorem}[Minimum sum AoI at the coordinating relay and destination nodes]\label{thm:min_AoI}
		For error-free symmetric IoT network, the minimum sum AoI  at the coordinating relay and at the destination nodes are expressed as follows.
		\begin{align}
			\nonumber \min_{\pi} \sum_{k=1}^K g_k^{\pi}(t) &= \left( 1 - \mathbbm{1}\left(t \geq t'+1 \right)\right) tK + \mathbbm{1}\left(t \geq t'+1 \right) t'K  \\
			&- \frac{t'(t'-1)S}{2} ,\\
			\nonumber \min_{\pi} \sum_{k=1}^K h_k^{\pi}(t) &= \left( 1 - \mathbbm{1}\left( t \geq t'' + 1 \right) \right) t K + \mathbbm{1}\left( t \geq t'' + 1 \right) t'' K \\
			& - \frac{(t''-1)(t''-2)U}{2},
		\end{align} where $t' = \left\lceil \frac{K}{S} \right\rceil$ and $t'' = \left\lceil \frac{K}{U} \right\rceil + 1$.
		For $t > t''$,
		\begin{align}
			\min_{\pi} \sum_{k=1}^K g_k^{\pi}(t) &= t'K  - \frac{t'(t'-1)S}{2} ,\label{eq:sumAoI_ap_large_t}\\
			\min_{\pi} \sum_{k=1}^K h_k^{\pi}(t) &= t'' K - \frac{(t''-1)(t''-2)U}{2}. \label{eq:sumAoI_des_large_t}
		\end{align}
	\end{theorem}
	\begin{IEEEproof}
		Since the greedy policy is optimal for minimizing the sum AoI for any $t$, we have,
		\begin{align}
			\min_{\pi} \sum_{k=1}^K g_k^{\pi}(t) &= \sum_{k=1}^K g_k^{G}(t), \\
			\min_{\pi} \sum_{k=1}^K h_k^{\pi}(t) &= \sum_{k=1}^K h_k^{G}(t).
		\end{align}
		By Lemma \ref{lem:sum_AoI} and Proposition \ref{prop:greedy_reduction}, the sum AoI at the coordinating relay can be written as
		\begin{align}
			\sum_{k=1}^K g_k^{G}(t) &= t K -\sum_{\tau=1}^{t-1} \min\left\lbrace \tau S , K \right\rbrace . \label{thm3_temp1}
		\end{align}
		Then \eqref{thm3_temp1} can be simplified as	
		\begin{align}
			\sum_{k=1}^K g_k^{G}(t) &= t K -\sum_{\tau=1}^{t-1} \min\left\lbrace \tau S , K \right\rbrace ,\\ 
			& = tK - \left( \sum_{\tau=1}^{t'-1} \tau S + \sum_{\tau=t'}^{t-1} K \right) \\
			& = tK - \frac{t'(t'-1)S}{2} - \mathbbm{1}(t \geq t'+1)(t -t')K, \\
			\nonumber & = \left( 1 - \mathbbm{1}\left(t \geq t'+1 \right)\right) tK + \mathbbm{1}\left(t \geq t'+1 \right) t'K  \\
			& - \frac{t'(t'-1)S}{2}.
		\end{align} where $t' = \left\lceil \frac{K}{S} \right\rceil$.
		With similar derivations,
		\begin{align}
			\sum_{k=1}^K h_k^{G}(t) & = t K - \sum_{\tau=1}^{t-1} \min\left\lbrace (\tau-1)U, K \right\rbrace , \\
			\nonumber & = \left( 1 - \mathbbm{1}\left( t \geq t'' + 1 \right) \right) t K + \mathbbm{1}\left( t \geq t'' + 1 \right) t'' K \\
			&- \frac{(t''-1)(t''-2)U}{2},
		\end{align} where $t'' = \left\lceil \frac{K}{U} \right\rceil + 1$.
	\end{IEEEproof}
	
	Since error-free symmetric IoT network can achieve lower minimum sum AoI than error-prone symmetric network, the result of Theorem \ref{thm:min_AoI} reveals the fundamental limit of AoI in a relay-aided symmetric IoT network. Theorem \ref{thm:min_AoI} also implicates that  for large $t$ such that all the sensors and destinations are chosen at least once, the sum AoI  at the coordinating relay and destination nodes does not increase but becomes a constant which is independent of $t$ since $t'$ and $t''$ are constant values. Furthermore, as shown in \eqref{eq:sumAoI_ap_large_t} and \eqref{eq:sumAoI_des_large_t}, both the sum AoIs at the coordinating relay and destination nodes are not functions of $t$ for $t>t''$. Thus, regardless of the value of $K$, $S$, and $U$, the sum AoI is not growing for large $t$.  This is counter-intuitive because, for a sensor network which monitors a large number of physical processes but has a restriction on  the number of processes to sample and update in each time slot, it is expected that the AoIs keep increasing as time goes on. However, Theorem \ref{thm:min_AoI} implies that if we sample and update optimally, any symmetric IoT sensor network can achieve constant sum AoI regardless of the number of sensors and destination nodes, and capability of sampling and updating in each time slot.

	\subsection{Optimal policy in an error-prone symmetric system}
	In the previous subsection, we have shown that the greedy policy minimizes the average sum AoI at the destination nodes for the error-free symmetric IoT network. In this subsection, by using stochastic dominance, we show that, in the presence of transmission errors caused by outage, the greedy policy also minimizes the average sum AoI at the destination nodes. Although the AoI evolution path of IoT sensor network becomes stochastic due to outage events,  the greedy policy is still optimal as we have proved under the deterministic evolution in the error-free scenario.
	
	\begin{theorem}[Optimality of the greedy algorithm for error-prone symmetric IoT network]\label{thm:opt_greedy_error}
		For symmetric network with identical outage probability and equal weight, $p_k = p \in \left[ 0, 1 \right)$, $q_k = q \in \left[ 0, 1 \right)$, and $w_k = \frac{1}{K}$, $\forall k$,  the greedy sampling and updating policy is optimal for minimizing the average sum AoI at the destination nodes.
		\begin{align}
			G = \arg\min_{\pi} \frac{1}{TK}\sum_{t=1}^T \mathbb{E} \left[ \sum_{k=1}^K h_k^{\pi}(t) \right].
		\end{align}
	\end{theorem}
	\begin{IEEEproof}
		We will use stochastic dominance to prove that the greedy policy is optimal in the error-prone network using the optimality of the greedy policy in the error-free network.
		\begin{definition}[Stochastic Dominance]
			For random variables $X$ and $Y$, $X$ is stochastically dominant over $Y$ if $\Pr \left[ f(X) > z \right]  \geq \Pr \left[ f(Y) > z \right]$, $\forall z \in \mathbb{R}$, $\forall f \in \mathcal{F}$, where $\mathcal{F}$ is the set of measurable functions which maps from the domain of $X$ and $Y$ to positive real value.
		\end{definition}
		In our problem, the time-average of sum AoI is of our interest. Thereby, let us treat the instantaneous sum AoI given from a policy as a random variable. If the instantaneous sum AoI of any policy is stochastically dominant over that of the greedy policy, we have $\Pr \left[ \frac{1}{T}\sum_{\tau=1}^T\sum_{k=1}^K h_k^{\pi}(\tau) > z \right] \geq \Pr \left[ \frac{1}{T}\sum_{\tau=1}^T\sum_{k=1}^K h_k^{G}(\tau) >z \right]$ for any $\pi$. Since we can express average as
		\begin{align}
			\mathbb{E} \left[ \frac{1}{T}\sum_{\tau=1}^T \sum_{k=1}^K h_k^{\pi}(\tau) \right] = \int_{0}^{\infty} \Pr \left[ \frac{1}{T}\sum_{\tau=1}^T\sum_{k=1}^K h_k^{\pi}(\tau) > z \right] dz,
		\end{align}	we can conclude that average sum AoI of the greedy policy is the minimum.\\	
		In \cite{Kadota_TNET2018, Stoyan_Book, Raghunathan_INFOCOM2008, Ganti_TIT2007, Bhattacharya_TAC1989}, for given two random variables, $X$ and $Y$, $X$ is stochastically dominant over $Y$ if there exist two stochastic processes $X(t)$ and $Y(t)$ which satisfy
		\begin{itemize}
			\item $X$ and $X(t)$ have the same probability distribution	
			\item $Y$ and $Y(t)$ have the same probability distribution
			\item $X(t)$ and $Y(t)$ are defined on a common probability space
			\item $X(t) \geq Y(t)$, with probability $1$ for any $t$.
		\end{itemize}	
		Now, we will prove that AoI of any policy is stochastically dominant over AoI of the greedy policy by showing there exist corresponding stochastic processes such that above conditions are met. First, consider stochastic process $X(t) = \sum_{k=1}^K h_k^{\pi}(t)$ and $Y(t) = \sum_{k=1}^K h_k^{G}(t)$ for an arbitrary policy $\pi$ and the greedy policy, respectively. Then, the randomness of $X(t)$ and $Y(t)$ comes from   wireless channels which cause   outages. Therefore, a set of all feasible combinations of outage occurrences in transmissions of the selected processes becomes the probability space for $X(t)$ and $Y(t)$. Then, since any policy chooses sensors or destination nodes without affecting wireless channels or communication links, an occurrence of outage which depends on   the wireless channel is independent of a policy. In other words, regardless of the selection of $\pi$ and the greedy policy, the occurrences of outage events are independently given. Hence, given that $\pi$ selects IoT sensor $k$ and the greedy policy selects IoT sensor $k'$, the probability that an outage occurs in the communication links with IoT sensor $k$ and $k'$ is the same. In this context, we can define a sample path which is evolution of AoI at the coordinating relay and at the destination nodes following a given policy over the entire time-horizon. Let us consider a sample path $\alpha^{\pi}$ determined by an arbitrary policy $\pi$. Due to the independence of outage events, there exists a corresponding sample path $\alpha^{G}$ following the greedy policy which have the same realization of outage events as $\alpha^{\pi}$. In other words, $\alpha^{\pi}$ can show different selection of sampling and updating from $\alpha^{G}$ but the outage realizations of all the communication links at each time becomes identical in the sample paths $\alpha^{\pi}$ and $\alpha^{G}$. Since the outage probability is the same for both $\pi$ and the greedy policy, $X(t)$ and $Y(t)$ have the common probability space. Thus, $X(t)$ and $Y(t)$ satisfy the first three conditions. Now, we will show the last condition, $X(t) \geq Y(t)$ with probability $1$ for any $t$. If an outage occurs in transmission, the corresponding AoI value increases by $1$ regardless of policy. Therefore, even if the decisions made by $\alpha^{\pi}$ and $\alpha^{G}$ for sampling and updating are different, the corresponding AoI value for the decisions will be the same for $\pi$ and the greedy policy when outages happen in both of $\alpha^{\pi}$ and $\alpha^{G}$. In other words, only the decisions with non-outage event make distinction between different policies. As a consequence, we can only consider sampling and updating where outage does not occur. When the effect of outage can be neglected, the error-prone symmetric IoT network becomes equivalent to the error-free symmetric case, which we already show that the Greedy policy minimizes the average sum AoI. Therefore, the greedy policy minimizes the average sum AoI for the error-prone symmetric IoT network.
	\end{IEEEproof}
	
	Theorem \ref{thm:opt_greedy_error} implicates that even in the existence of errors, if outage occurs independently and identically across the sensors and destination nodes, the greedy algorithm is still optimal.
	
	\subsection{General IoT Network}
	This subsection discusses a general asymmetric IoT sensor network, which involves arbitrary weights, the number of sensors to sample, and the number of destination nodes to update. The order of the weighted AoI gap varies depending on the weights' values, making it impossible to quantify the reduction of the AoI gap analytically. Consequently, we propose a deep reinforcement learning approach based on the Markov decision process (MDP) to minimize the average weighted sum of AoI.
	
	To design MDP, we need to define states, actions, and rewards. First, a state can be defined as a concatenation of the AoIs at the coordinating relay and at the destination. Accordingly, a state $x(t)$ at time slot $t$ is represented as
	\begin{align}
		x(t) = \left( \left\lbrace g_k(t) \right\rbrace_{k=1}^K, \left\lbrace h_k(t) \right\rbrace_{k=1}^K \right).
	\end{align}
	An action is the selection of sensors for sampling and destination nodes for updating by the coordinating relay. An action at time slot $t$ denoted as $\alpha(t)$ becomes
	\begin{align}
		\alpha(t) = \left( \mathcal{S}(t), \mathcal{U}(t) \right).
	\end{align}
	A policy is defined as a mapping from state $x(t)$ to an action $\alpha(t)$. Namely, $\pi(x(t)) = \alpha(t)$.\\ 
	We define a reward as the weighted sum of AoI difference from the current state and the next states.  Hence, the accumulated reduction is to be maximized, which results in the minimum average weighted sum AoI. In other words, the reward obtained by action $\alpha(t)$ in state $x(t)$, $r\left(x(t), \alpha(t)\right)$, becomes
	\begin{align}
		r(x(t), \alpha(t)) = \sum_{k=1}^K -w_k \left( h_k(t+1) - h_k(t) \right).
	\end{align}
	Based on this formulation, we can solve the following MDP problem which finds a policy achieving near-optimal performance.
	\begin{align}
		&\max_{\pi} \sum_{t=1}^T \gamma^t r\left(x(t), \pi(x(t)) \right), \\
		\text{s.t. } &\left| \mathcal{S}(t) \right| = S , \hspace{5pt} \forall t, \\
		& \left| \mathcal{U}(t) \right| = U, \hspace{5pt} \forall t,
	\end{align} where $\gamma$ is a discount factor.\\
	To solve the MDP problem, we utilize a Q-learning algorithm where the state-action value denoted as $Q(x(t), \alpha(t))$ is updated as follows.
	\begin{align}
		&Q^{i+1}(x(t), \alpha(t)) = Q^{i}(x(t), \alpha(t)) \\
		\nonumber  &+ \eta \left( r(t) + \gamma \max_{\alpha(t+1)} Q^{i}(x(t+1), \alpha(t+1)) - Q^{i}(x(t), \alpha(t)) \right) .
	\end{align}
	Thus, in the $i$-th iteration, we need to have values of $Q^{i}(x(t), \alpha(t))$ for all $x(t)$ and $\alpha(t)$. However, in our problem, the state space and action space are huge. Thus, Q-learning which requires a table of which size is product of cardinality of state space and action space cannot be applied to our problem. Replacing the table with a neural network, Deep Q-learning is more favorable.
	
	In deep Q-learning, we define a Deep Q-network (DQN) which approximates the Q-value for each state and action. In the training period, we collect states, actions, rewards,  and the next states that the DQN experiences. By utilizing those collected data, we train the DQN to approximate the Q-values with small gap by updating parameters of DQN. To optimize parameters of DQN, we need to define the loss function which shows how far our neural network is from the true Q-value. In our problem, we use the smooth $l1$-loss function. 
	
	The parameters of DQN are updated toward the direction of reducing the loss between the DQN value at the current state and the sum of the reward and the DQN value at the next state. The procedure from collecting data and updating neural network is repeated until the neural network converges to the true Q-values. After the DQN converges to the true Q-values, we use the trained neural network for a DQN policy. The DQN policy is given as 
	\begin{align}
		\pi^{\text{DQN}}(x(t)) = \argmax_{a(t)} Q(x(t), a(t)).
	\end{align}
	In our problem, the state space and action space are exponentially scaled with respect to $K$, $S$, and $U$. Unfortunately, due to the large size of these spaces, the convergence of the DQN policy is significantly time-consuming and may pose a challenge of converging to the global minima.
	
	\section{Numerical Results}\label{sec:simulations}
	In this section, we present the performance evaluation of different policies aiming at minimizing the average weighted sum of AoI at the destination nodes. We compare the average weighted sum AoI for several policies in various environments, including the greedy policy, DQN policy, random policy \cite{Kadota_TNET2018}, threshold policy \cite{Klügel_INFOCOM2019}, MAF policy \cite{Sun_INfocom2018}, and index policy \cite{Kadota_Allerton2016}. The random policy selects a feasible sampling and updating set with a uniform probability distribution. The threshold policy selects nodes whose AoIs exceed a predefined threshold value. The MAF policy selects nodes with the highest AoI values. By leveraging the MDP formulation, the index policy computes Whittle's indexes for each node as an initial step. In our system setting, the index for sensor $k$ at time $t$ is given as
		\begin{align}
			C(t) = \frac{w_k}{2}(1 -  p_k) g_k^{\pi}(t) \left(  g_k^{\pi}(t) + \frac{1 + p_k}{1 - p_k} \right)    ,  \label{eq:index}
		\end{align} where $g_k^{\pi}(t)$ is the AoI at the relay for sensor $k$ at time $t$ with a policy $\pi$. Similarly, the indexes for destination nodes can be obtained by replacing the corresponding parameters. Once we have computed all the indexes for both sensors and destination nodes, we can schedule the nodes with higher indexes for updating.
	
		\begin{figure}[t!]
			\centering
			\includegraphics[scale=0.6]{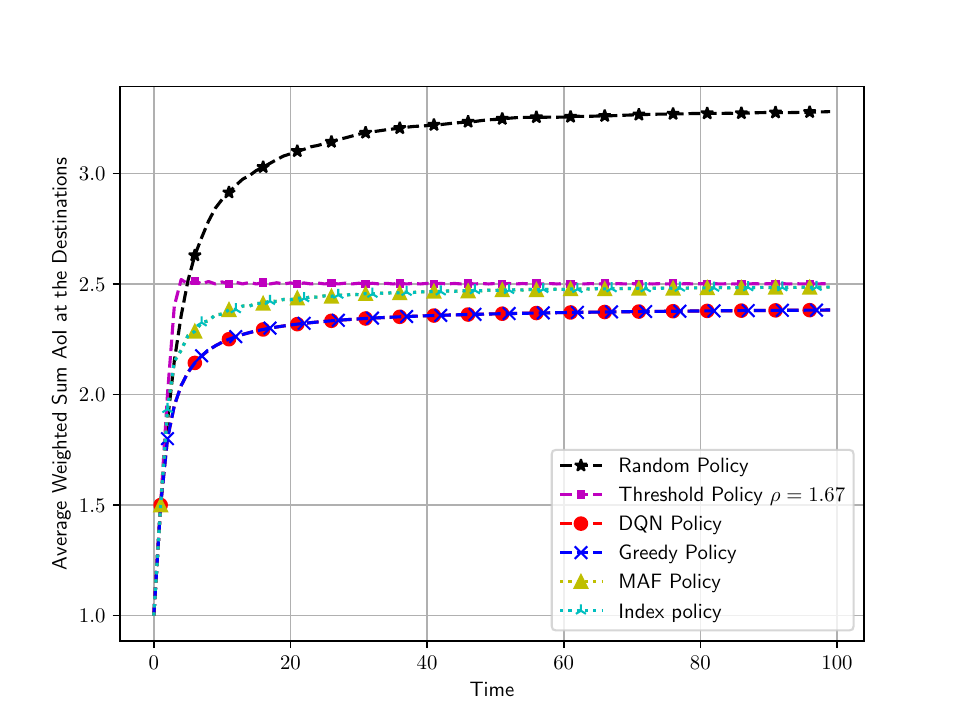}
			\caption{Comparison of the average weighted sum AoIs at the destination nodes in an error-free symmetric IoT network for $T=100$}\label{fig:T_20}
		\end{figure}
	
		First, we present the sum of AoI for different policies in an error-free symmetric IoT network. For this analysis, we consider the following parameters: $T = 100$, $K=5$, $S=3$, $U=3$, $w_k=1/K$, and $p_k=q_k =0$ for all $k$.   Fig. \ref{fig:T_20} illustrates the average weighted sum AoI at the destination nodes for the different policies.
		As stated in Theorem \ref{thm:min_AoI}, the sum of AoI at the coordinating relay and the destination nodes remains constant after a certain time slot for the greedy policy. In this case, the performance of the greedy policy, which is proven to be optimal, is equivalent to that of the DQN policy. 
		
		The threshold policies with $\rho = \frac{S}{K} = 1.67$ exhibit higher average weighted sum AoI values in the earlier time slots. Since the pure threshold policies cannot always satisfy the constraint, there are instances where the threshold policy may select fewer sensors or destination nodes than $S$ or $U$. Conversely, for low threshold values, the pure threshold policies may choose a larger number of nodes. Therefore, an additional step is necessary for the pure threshold policies to satisfy the constraint. If the pure threshold policy selects more nodes than the allowed number, we need to discard some of the selected nodes. Ideally, we prioritize keeping nodes with higher AoI values since our objective is to minimize AoI. However, when using a threshold policy with a high threshold value, the constraint may not be satisfied exactly, resulting in a lower number of selected nodes than the maximum allowed number. Consequently, the performance of the threshold policy with a high threshold value is strictly suboptimal. 
		
		Conversely, the MAF policy behaves identically to the greedy policy when sampling new sensory data from sensors since both policies select the $S$ sensors with the highest AoI. However, in terms of updating, the MAF policy solely considers the AoI value at the destination nodes. As a result, the reduction in AoI through updating may not be significant when the AoI at the relay node is relatively high.
		
		Additionally, if we consider the index as a simplified quadratic function of the AoI value, as shown in Equation \eqref{eq:index}, the index policy for an error-free symmetric network selects nodes based on their higher AoI values, which is essentially equivalent to the MAF policy \cite{Kadota_Allerton2016}. Therefore, the performance of the MAF policy and the index policy is identical.
		\begin{figure}
			\centering
			\includegraphics[scale=0.6]{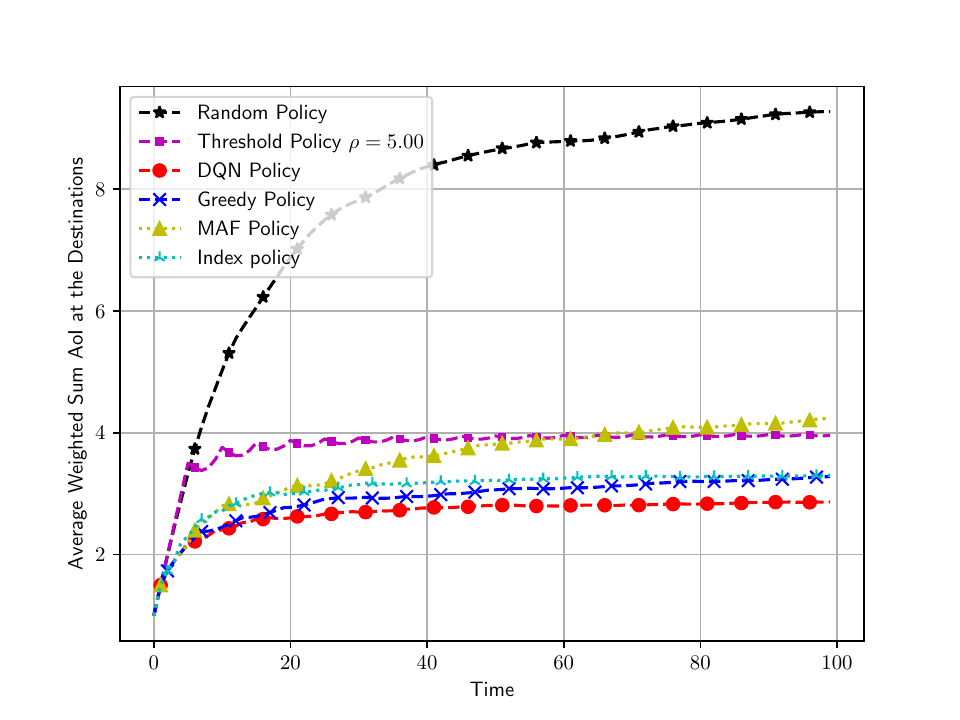}
			\caption{Comparison of the average weighted sum AoIs at the destination nodes in an error-free asymmetric IoT network for $\eta = 0.2$} \label{fig_asymm_02}
		\end{figure}
	
		Moreover, we consider error-free asymmetric network. To investigate asymmetricity of weights, we use geometric weights defined as
		\begin{align}
			w_{k+1} = \eta w_k, \label{eq:g_weight}
		\end{align} where $\eta$ is a common ratio and $w_1$ is determined to satisfy $\sum_{k=1}^K w_k = 1$.
		We compare the average weighted sum of AoI for the following parameters: $K=5$, $T=20$, $S=U=1$, $p_k=q_k=0$, and $\eta=0.2$.  Fig. \ref{fig_asymm_02} illustrates the results.
		The DQN policy exhibits the lowest average weighted sum AoI at the destination nodes. This can be attributed to the presence of asymmetric weights in the network. However, the performance gap between the greedy policy and the DQN policy is low. Considering the significant training cost associated with the DQN policy due to its exponentially large state and action spaces, the greedy policy can be a favorable alternative, even in an asymmetric network. In this scenario, where the weights are different, the index policy outperforms the MAF policy. 
		\begin{figure}[t!]
			\centering
			\includegraphics[scale=0.6]{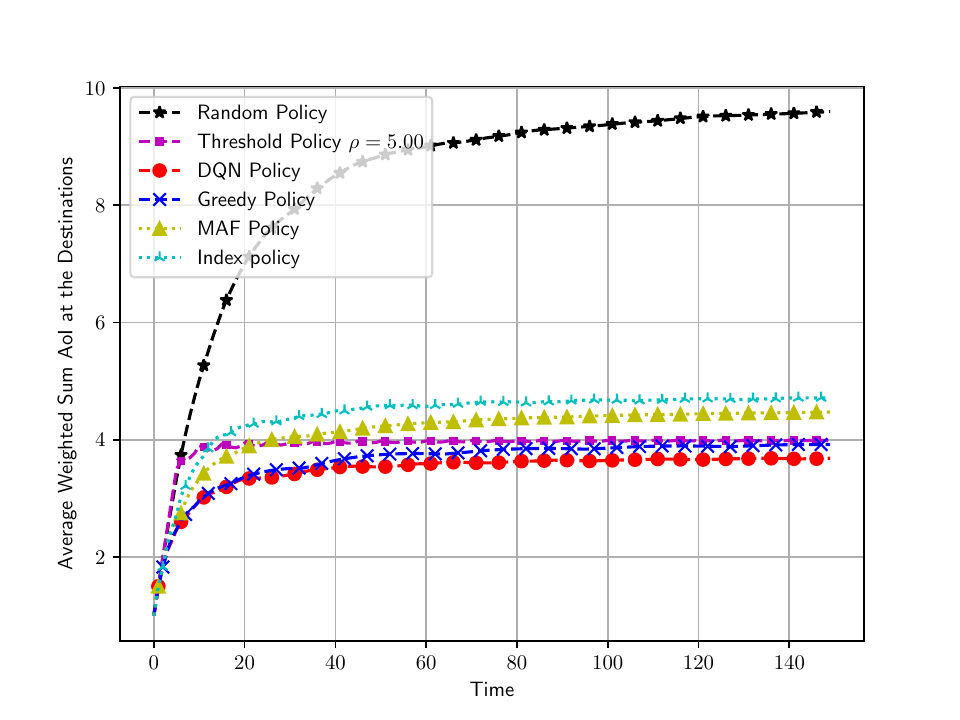}
			\caption{Comparison of the average sum AoI at the destination node in an error-free asymmetric IoT networks for $\eta = 0.5$}\label{fig_asymm_05}
		\end{figure}
		If we increase $\eta$ to $0.5$, the average weighted sum of AoI at the destination nodes for different policies is depicted in  Fig. \ref{fig_asymm_05}. When $\eta = 0.5$, the weight vectors are given as $\mathbf{w} = \left(0.5161, 0.2581, 0.129, 0.0645, 0.0323\right)$. Consequently, destination node $1$ contributes to more than half of the overall performance. In this highly biased scenario, the DQN policy demonstrates a low average weighted sum of AoI.	
	
		\begin{figure}[h!]
			\centering
			\includegraphics[scale=0.6]{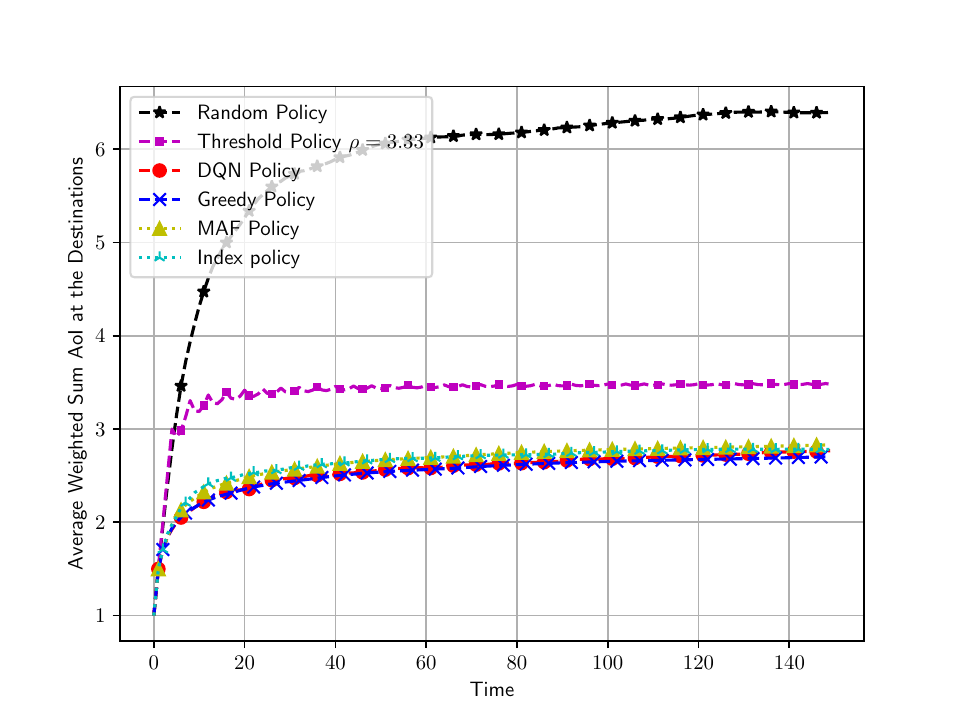}
			\caption{Comparison of the average sum AoI at the destination node in an error-free asymmetric IoT networks for $K=10, S=U=3$} \label{fig_K10_SU3}
		\end{figure}
	
		\begin{figure}[h!]
			\centering
			\includegraphics[scale=0.6]{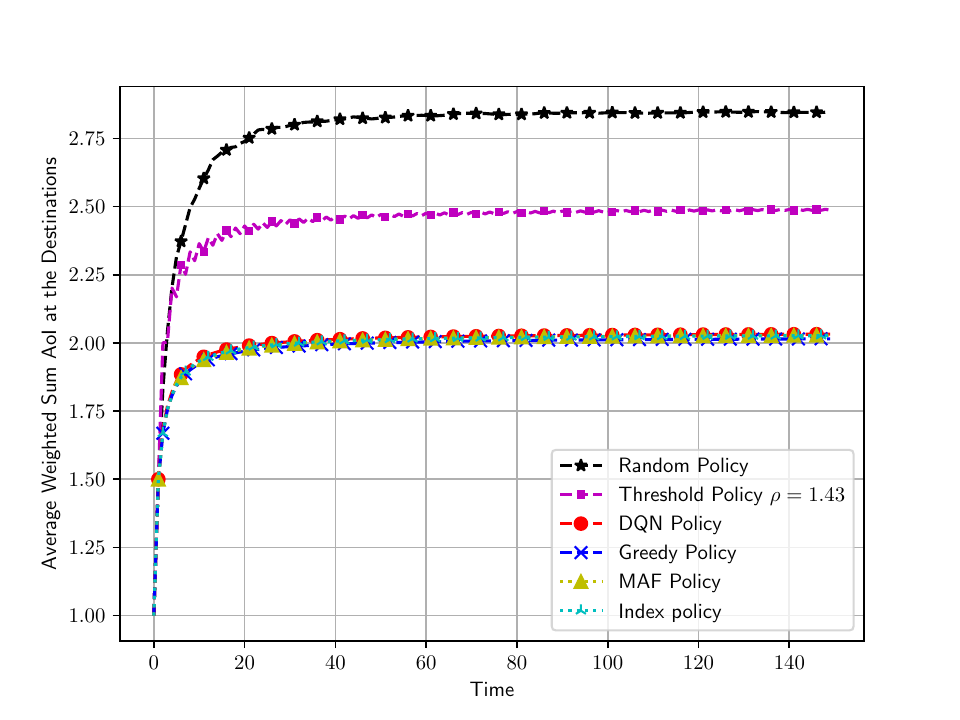}
			\caption{Comparison of the average sum AoI at the destination node in an error-free asymmetric IoT networks for $K=10, S=U=7$} \label{fig_K10_SU7}
		\end{figure}
	
		We further compare the average weighted sum of AoI for different policies, considering different numbers of sensors and destination nodes to be updated. In Fig. 5, we examine the case when $K=10$, $S=U=3$, and $\eta=0.5$, showcasing the performance of different policies. Furthermore, in Fig. \ref{fig_K10_SU7}, we present the average weighted sum AoI by increasing $S=U=7$. As the number of selected sensors and destination nodes increases, the performance gap between the different policies diminishes. For example, when $K=10$ and $S=U=7$, at least $4$ destination nodes are commonly selected by different policies. This implies that as $S$ and $U$ grow, the overlap in selected nodes among the policies also increases, resulting in reduced differences in performance. Additionally, in an asymmetric network with $\eta=0.5$, the resulting weight vector is $\mathbf{w} = (0.5005, 0.2502, 0.1251, 0.0626, 0.0313, 0.0156, \\ 0.0078, 0.0039, 0.0020, 0.0010)$. In this case, only a few destination nodes (from 1 to 4) contribute to over 93\% of the overall performance. Consequently, the performance gap between policies becomes smaller, emphasizing the importance of selecting an appropriate scheduling policy in a symmetric network with limited wireless resources.
		
		\begin{figure}[h!]
			\centering
			\includegraphics[scale=0.6]{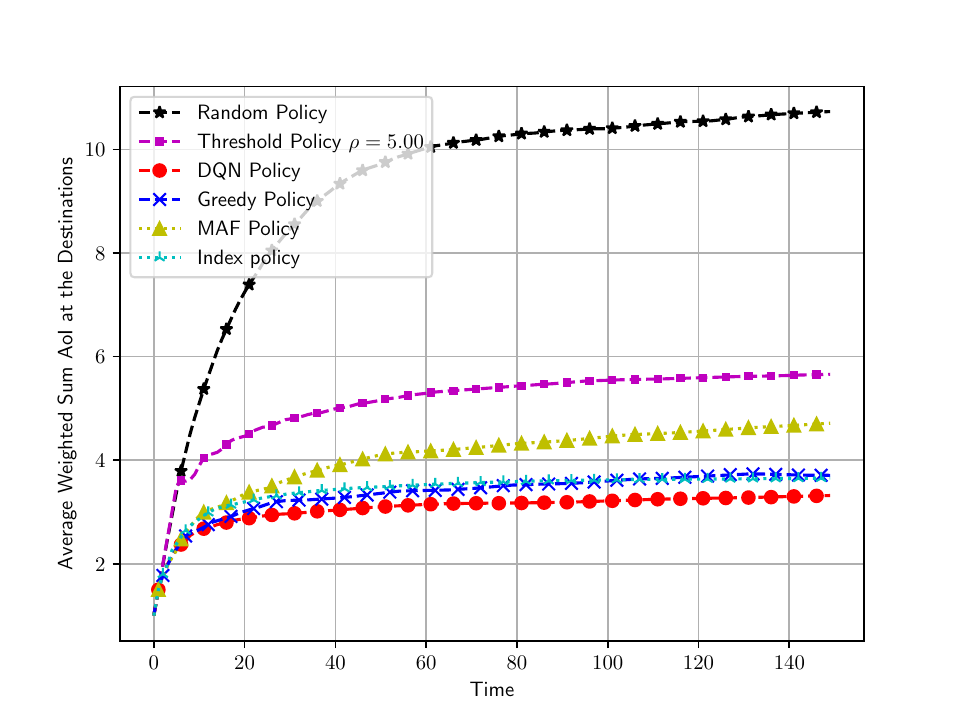}
			\caption{Comparison of the average sum AoI at the destination node in an error-prone asymmetric IoT networks for $p_k = q_k = 0.1, \hspace{5pt} \forall k$} \label{fig_err_01}
		\end{figure}
	
		In an error-prone asymmetric network, we compare the average weighted sum of AoI for different policies in the Fig. \ref{fig_err_01}, assuming $p_k = q_k = 0.1$ for all $k$. In this scenario, the DQN policy achieves the lowest average weighted sum AoI. Furthermore, the index policy, which takes into account the error probability when calculating the index, exhibits comparable performance. Surprisingly, the greedy policy, which neither requires training nor information regarding the error distribution, exhibited similar results. Consequently, the greedy policy presents itself as a promising solution for minimizing AoI while maintaining a low implementation cost.
	
	\section{Conclusion}\label{sec:conclusion}
	In this paper, we have considered an IoT sensor network where sensors deliver sampled data of physical processes to destination nodes via a relay which coordinates sampling and updating. When the number of communication links simultaneously activated is limited, the coordinating relay needs to choose sensors to sample fresh data and destination nodes to be updated carefully. We have investigated the efficient policy for the decision of the coordinating relay. When communication links are highly stable and the importance of physical process is identical, the necessary and sufficient conditions for optimal policy have been figured out. In addition to that, the greedy policy has been proved to satisfy those conditions. Furthermore, the expression for the minimum sum AoI has been obtained. Even in the case that communication errors are considered, the greedy policy has been  proved to be  still optimal. When we consider a general IoT network with different importance of physical processes with non-negligible communication error, we have proposed a DQN policy. Our  numerical results have shown that the greedy policy can show similar performance to the DQN policy. Therefore, when the training cost is not affordable, the greedy policy can be an alternative to achieve low average weighted sum AoI at the destination nodes.
	
	\ifCLASSOPTIONcaptionsoff
	\newpage
	\fi


\begin{thebibliography}{1}
		
		\bibitem{Yates_Arxiv2020}
		R. D. Yates, Y. Sun, D. R. Brown, S. K. Kaul, E. Modiano and S. Ulukus, "Age of Information: An Introduction and Survey," \textit{IEEE J. Sel. Areas Commun.,} vol. 39, no. 5, pp. 1183-1210, May 2021.
		
		\bibitem{Kosta_FTN2017}
		A. Kosta, N. Pappas, and V. Angelakis, "Age of information : A new concept, metric, and tool," in \textit{Foundations and Trends in Networking,} vol. 12, no. 3, pp. 162-259, 2017.
		
		\bibitem{Kaul_ICSCSM2011}
		S. Kaul, M. Gruteser, V. Rai and J. Kenney, "Minimizing age of information in vehicular networks," \textit{Proc. IEEE Communications Society Conference on Sensor, Mesh and Ad Hoc Communications and Networks}, 2011, pp. 350-358.
		
		\bibitem{Klügel_INFOCOM2019}
		M. Klugel, M. H. Mamduhi, S. Hirche and W. Kellerer, "AoI-penalty minimization for networked control systems with packet loss," \textit{Proc. IEEE Conference on Computer Communications Workshops (INFOCOM WKSHPS)}, 2019, pp. 189-196.
		
		\bibitem{Krikidis_WCL2019}
		I. Krikidis, "Average age of information in wireless powered sensor networks," \textit{IEEE Wireless Commun. Letters,} vol. 8, no. 2, pp. 628-631, April 2019.
		
		\bibitem{Tsai_INFOCOM2020}
		C. Tsai and C. Wang, "Unifying AoI minimization and remote estimation - optimal sensor/controller coordination with random two-way delay,"  \textit{Proc. IEEE Conference on Computer Communications (INFOCOM)}, Canada, 2020, pp. 466-475.
		
		\bibitem{Arafa_TIT2020}
		A. Arafa, J. Yang, S. Ulukus and H. V. Poor, "Age-minimal transmission for energy harvesting sensors with finite batteries: online policies,"  \textit{IEEE Trans. Inf. Theory,} vol. 66, no. 1, pp. 534-556, Jan. 2020.
		
		\bibitem{Kaul_SMACN2011}
		S. Kaul, M. Gruteser, V. Rai and J. Kenney, "Minimizing age of information in vehicular networks," \textit{Proc. IEEE Communications Society Conference on Sensor, Mesh and Ad Hoc Communications and Networks,} 2011, pp. 350-358.
		
		\bibitem{Zhang_TWC2021}
		S. Zhang, L. Wang, H. Luo, X. Ma and S. Zhou, "AoI-Delay Tradeoff in Mobile Edge Caching With Freshness-Aware Content Refreshing,"  \textit{IEEE Trans. Wireless Commun.,} vol. 20, no. 8, pp. 5329-5342, Aug. 2021.
		
		\bibitem{Hou_TWC2005}
		Y. T. Hou, Yi Shi, H. D. Sherali and S. F. Midkiff, "On energy provisioning and relay node placement for wireless sensor networks," \textit{IEEE Trans. Wireless Commun.,} vol. 4, no. 5, pp. 2579-2590, Sept. 2005.
		
		\bibitem{Lloyd_TCOMP2007}
		E. L. Lloyd and G. Xue, "Relay node placement in wireless sensor networks," \textit{IEEE Trans. on Comput.}, vol. 56, no. 1, pp. 134-138, Jan. 2007.
		
		\bibitem{Sisinni_TII2018}
		E. Sisinni, A. Saifullah, S. Han, U. Jennehag, and M. Gidlund, "Industrial Internet of Things: Challenges, opportunities, and directions," \textit{IEEE Trans. Ind. Informat.}, vol. 14, no. 11, pp. 4724-4734, Nov. 2018.
			
		\bibitem{Zakeri_Arxiv2022}
		A. Zakeri, M. Moltafet, M. Leinonen, and M. Codreanu, "Dynamic Scheduling for Minimizing AoI in Resource-Constrained Multi-Source Relaying Systems with Stochastic Arrivals." in \textit{arXiv} preprint arXiv:2203.05656, 2022.
		
		\bibitem{Kaul_INFOCOM2012}
		S. Kaul, R. Yates and M. Gruteser, "Real-time status: How often should one update?," \textit{Proc. IEEE Conference on Computer Communications (INFOCOM)}, 2012, pp. 2731-2735.
		
		\bibitem{Yates_TIT2019}
		R. D. Yates and S. K. Kaul, "The age of information: Real-time status updating by multiple sources," \textit{IEEE Trans. Inf. Theory,} vol. 65, no. 3, pp. 1807-1827, March 2019.
		
		\bibitem{Hsu_TMC2019}
		Y. -P. Hsu, E. Modiano and L. Duan, "Scheduling Algorithms for Minimizing Age of Information in Wireless Broadcast Networks with Random Arrivals,"  \textit{IEEE Trans. Mobile Comput.,} vol. 19, no. 12, pp. 2903-2915, Dec. 2020.
		
		\bibitem{Kalør_WCL2019}
		A. E. Kalor and P. Popovski, "Minimizing the age of information from sensors with common observations,"  \textit{IEEE Wireless Commun. Letters}, vol. 8, no. 5, pp. 1390-1393, Oct. 2019.
		
		\bibitem{Kosta_JCN2019}
		A. Kosta, N. Pappas, A. Ephremides and V. Angelakis, "Age of information performance of multiaccess strategies with packet management,"  \textit{Journal of Commun. and Netw.,} vol. 21, no. 3, pp. 244-255, June 2019.
		
		\bibitem{Sun_TCOM2020}
		J. Sun, Z. Jiang, B. Krishnamachari, S. Zhou and Z. Niu, "Closed-form Whittle's index-enabled random access for timely status update,"  \textit{IEEE Trans. Commun.,} vol. 68, no. 3, pp. 1538-1551, March 2020.
		
		\bibitem{Javani_Globecom2019}
		A. Javani, M. Zorgui and Z. Wang, "Age of information in multiple sensing,"  \textit{Proc. IEEE Global Communications Conference (GLOBECOM)}, Waikoloa, 2019, pp. 1-6.
		
		\bibitem{Bastopcu_TWC2021}
		M. Bastopcu and S. Ulukus, "Information Freshness in Cache Updating Systems," \textit{IEEE Trans. Wireless Commun.,} vol. 20, no. 3, pp. 1861-1874, March 2021.
		
		\bibitem{Chen_Arxiv2020}
		H. Chen, Y. Gu and S. -C. Liew, "Age-of-Information Dependent Random Access for Massive IoT Networks," \textit{Proc. IEEE Conference on Computer Communications Workshops (INFOCOM WKSHPS)}, 2020, pp. 930-935.
		
		\bibitem{Chen_Arxiv2019}
		X. Chen, K. Gatsis, H. Hassani and S. S. Bidokhti, "Age of Information in Random Access Channels," \textit{Proc. IEEE International Symposium on Information Theory (ISIT)}, 2020, pp. 1770-1775.
		
		\bibitem{Ceran_INFOCOM2019}
		E. T. Ceran, D. Gunduz and A. Gyorgy, "Reinforcement learning to minimize age of information with an energy harvesting sensor with HARQ and sensing Cost,"  \textit{Proc. IEEE Conference on Computer Communications Workshops (INFOCOM WKSHPS)}, 2019, pp. 656-661.
		
		\bibitem{Abd_TCOM2020}
		M. A. Abd-Elmagid, H. S. Dhillon and N. Pappas, "A reinforcement learning framework for optimizing age of information in RF-powered communication systems," \textit{IEEE Trans. Commun.,} vol. 68, no. 8, pp. 4747-4760, Aug. 2020.
		
		\bibitem{Kadota_TNET2018}
		I. Kadota, A. Sinha, E. Uysal-Biyikoglu, R. Singh and E. Modiano, "Scheduling policies for minimizing age of information in broadcast wireless networks,"  \textit{IEEE/ACM Trans. Netw.}, vol. 26, no. 6, pp. 2637-2650, Dec. 2018.
		
		\bibitem{Zhou_TCOM2019}
		B. Zhou and W. Saad, "Joint status sampling and updating for minimizing age of information in the internet of things,"  \textit{IEEE Trans. Commun.,} vol. 67, no. 11, pp. 7468-7482, Nov. 2019. 
		
		
		
		
		\bibitem{Simiscuka_IWCMC2018}
		A. A. Simiscuka and G. Muntean, "Age of information as a QoS metric in a relay-based IoT mobility solution,"  \textit{Proc. International Wireless Communications \& Mobile Computing Conference (IWCMC)}, 2018, pp. 868-873.
		
		\bibitem{Cao_ICC2020}
		A. Cao, C. Shen, J. Zong and T. Chang, "Peak age-of-information minimization of UAV-aided relay transmission,"  \textit{Proc. IEEE International Conference on Communications Workshops (ICC Workshops)}, 2020, pp. 1-6.
		
		\bibitem{Zhou_WCSP2019}
		C. Zhou et al., "Deep RL-based trajectory planning for AoI minimization in UAV-assisted IoT,"  \textit{Proc. IEEE International Conference on Wireless Communications and Signal Processing (WCSP)}, 2019, pp. 1-6.
		
		\bibitem{Hu_IOTJ2020}
		H. Hu, K. Xiong, G. Qu, Q. Ni, P. Fan and K. B. Letaief, "AoI-minimal trajectory planning and data collection in UAV-assisted wireless powered IoT networks," in \textit{IEEE Internet of Things Journal}, vol. 8, no. 2, pp. 1211-1223, 15 Jan.15, 2021.
		
		\bibitem{Talak_Allerton2017}
		R. Talak, S. Karaman and E. Modiano, "Minimizing age-of-information in multi-hop wireless networks," \textit{Proc. IEEE Annual Allerton Conference on Communication, Control, and Computing (Allerton)}, Monticello, IL, 2017, pp. 486-493.
		
		\bibitem{Moradian_WCNC2020}
		M. Moradian and A. Dadlani, "Age of information in scheduled wireless relay networks," \textit{Proc. IEEE Wireless Communications and Networking Conference (WCNC)}, 2020, pp. 1-6.
		
		\bibitem{Maatouk_ITW2018}
		A. Maatouk, M. Assaad and A. Ephremides, "The age of updates in a simple relay network," \textit{Proc. IEEE Information Theory Workshop (ITW)}, 2018, pp. 1-5.
		
		\bibitem{Arafa_TWC2019}
		A. Arafa and S. Ulukus, "Timely Updates in Energy Harvesting Two-Hop Networks: Offline and Online Policies," \textit{IEEE Trans. Wireless Commun.}, vol. 18, no. 8, pp. 4017-4030, Aug. 2019.
		
		\bibitem{Sun_INfocom2018}
		Y. Sun, E. Uysal-Biyikoglu, and S. Kompella, "Age-optimal updates of multiple information flows," \textit{Proc. of IEEE Conference on Computer Communications Workshops (INFOCOM WKSHPS)}, 2018, pp. 136-141.
		
		\bibitem{Beytur_BLACK}
		H. B. Beytur and E. Uysal-Biyikoglu, "Minimizing age of information for multiple flows," \textit{Proc. of IEEE International Black Sea Conference on Communications and Networking (BlackSeaCom)}, 2018, pp. 1-5.
		
		\bibitem{Farazi_ISIT2018}
		S. Farazi, A. G. Klein and D. R. Brown, "Age of information in energy harvesting status update systems: When to preempt in service?," \textit{Proc. IEEE International Symposium on Information Theory (ISIT)}, 2018, pp. 2436-2440.
		
		\bibitem{Yi_INFO2020}
		M. Yi, X. Wang, J. Liu, Y. Zhang and B. Bai, "Deep reinforcement learning for fresh data collection in UAV-assisted IoT networks,"  \textit{Proc. IEEE Conference on Computer Communications Workshops (INFOCOM WKSHPS)}, 2020, pp. 716-721.
		
		\bibitem{Stoyan_Book}
		D. Stoyan, \textit{Comparison methods for queues and other stochastic models}, Wiley, 1983.
		
		\bibitem{Bhattacharya_TAC1989}
		P. P. Bhattacharya and A. Ephremides, "Optimal scheduling with strict deadlines," \textit{IEEE Trans. Autom. Control}, vol. 34, no. 7, pp. 721-728, July 1989.
		
		\bibitem{Raghunathan_INFOCOM2008}
		V. Raghunathan, V. Borkar, M. Cao and P. R. Kumar, "Index policies for real-time multicast scheduling for wireless broadcast systems,"  \textit{Proc. IEEE Conference on Computer Communications (INFOCOM)}, 2008, pp. 1570-1578.
		
		\bibitem{Ganti_TIT2007}
		A. Ganti, E. Modiano and J. N. Tsitsiklis, "Optimal transmission scheduling in symmetric communication models with intermittent connectivity," in \textit{IEEE Trans. Inf. Theory}, vol. 53, no. 3, pp. 998-1008, March 2007.
		
		\bibitem{Christos_Combopt}
		C. H. Papadimitriou, K. Steiglitz, Combinatorial optimization: algorithms and complexity. Courier Corporation, 1998.
		
		\bibitem{Kadota_Allerton2016}
		I. Kadota, E. Uysal-Biyikoglu, R. Singh, and E. Modiano, "Minimizing the age of information in broadcast wireless networks," \textit{Proc. IEEE Allerton Conference on Communication, Control, and Computing (Allerton)}, 2016, pp. 844-851. 
	
		
	\end{thebibliography}
\end{document}